\documentclass[pdflatex,sn-mathphys-num]{sn-jnl}% Math and Physical Sciences Numbered Reference Style

\usepackage{graphicx}
\usepackage{amsmath}
\usepackage{amssymb}
\usepackage[table]{xcolor}
\usepackage{algpseudocode}
\usepackage{algorithm}
\usepackage{newtxtext,newtxmath}
\usepackage{booktabs}
\usepackage{wrapfig}
\usepackage{subcaption}
\usepackage{tikz}
\usepackage{forloop}
\usepackage{multirow}
\usepackage{longtable}
\usepackage[most]{tcolorbox} % for \greenbox

\title{Mixed-precision numerics in scientific applications: survey and perspectives}
\author*[1]{Aditya Kashi} \email{kashia@ornl.gov}
\author[1]{Hao Lu} \email{luh1@ornl.gov}
\author[1]{Wesley Brewer} \email{brewerwh@ornl.gov}
\author[1]{David Rogers} \email{rogersdm@ornl.gov}
\author[1]{Michael Matheson} \email{mathesonma@ornl.gov}
\author[1]{Mallikarjun Shankar} \email{shankarm@ornl.gov}
\author[1]{Feiyi Wang} \email{fwang2@ornl.gov}
\affil*[1]{\orgdiv{National Center for Computational Sciences}, \orgname{Oak Ridge National Laboratory}, \orgaddress{\city{Oak Ridge}, \state{TN}, \postcode{37830}, \country{USA}}}

\let\bld\boldsymbol

\newtcolorbox{greenbox}{
breakable, enhanced, boxrule=0pt,frame hidden, borderline west={4pt}{0pt}{green!75!black}, colback=green!10!white, sharp corners
}

\newcommand*{\priority}[1]{
\begin{tikzpicture}[scale=0.15]%
    \draw (0,0.2) circle (0.75); %%For the alignment purpose
    \fill[fill opacity=0.5,fill=blue] (0,0) -- (90:1) arc (90:90-#1*3.6:0.75) -- cycle;
    \end{tikzpicture}
}

\newcommand*{\prioritytwo}[1]{
\begin{tikzpicture}[scale=0.15]%
    \draw (0,0.2) circle (0.75); %%For the alignment purpose
    \draw[fill opacity=0.5,fill=blue] (0,0)-- (90:1) arc (90:270:0.75) -- cycle ;
    \end{tikzpicture}
}

\newcounter{themenumber}

\newcommand*{\supportss}[3]{
    \forloop{themenumber}{0}{\value{themenumber} <#1}
    {%
        \priority{100}\hspace{-3px}
    }%
    \forloop{themenumber}{0}{\value{themenumber} <#2}
    {%
        \prioritytwo{50}\hspace{-3px}
    }%
    \forloop{themenumber}{0}{\value{themenumber} <#3}
    {%
        \priority{0}\hspace{-3px}
    }%
 }

\begin{document}

\abstract{%
%We observe three emerging trends in floating-point compute capabilities for science driven by advancements in high-performance, accelerator-based chip design and development. First,
The explosive demand for artificial intelligence (AI) workloads has led to a significant increase in silicon area dedicated to lower-precision computations on recent high-performance computing hardware designs. However, mixed-precision capabilities, which can achieve performance improvements of 8$\times$ compared to double-precision in extreme compute-intensive workloads, remain largely untapped in most scientific applications.
A growing number of efforts have shown that mixed-precision algorithmic innovations can deliver superior performance without sacrificing accuracy.
These developments should prompt computational scientists to seriously consider whether their scientific modeling and simulation applications could benefit from the acceleration offered by new hardware and mixed-precision algorithms.

In this survey, we (1) review progress across diverse scientific domains—including fluid dynamics, weather and climate, quantum chemistry, and computational genomics -- that have begun adopting mixed-precision strategies; (2) examine state-of-the-art algorithmic techniques such as iterative refinement, splitting and emulation schemes, and adaptive precision solvers; (3) assess their implications for accuracy, performance, and resource utilization; and (4) survey the emerging software ecosystem that enables mixed-precision methods at scale.
We conclude with perspectives and recommendations on cross-cutting opportunities, domain-specific challenges, and the role of co-design between application scientists, numerical analysts and computer scientists. Collectively, this survey underscores that mixed-precision numerics can reshape computational science by aligning algorithms with the evolving landscape of hardware capabilities.
 }


%\runninghead{Mixed precision in science}

\maketitle

\tableofcontents

\section{Introduction}

%In numerical computing, two types of errors generally arise: \textit{round-off} error and \textit{truncation} error. Round-off error occurs because only a finite number of unique real numbers can be represented on a digital computer using a given format with a certain number of bits.
%The `gap' between the number 1 and the smallest number after it that can be represented in a number format is called its `machine epsilon'.
%On the other hand, truncation error is not related to machine precision. It arises when computations based on infinite sequences and series are truncated after a certain number of terms in a realistic algorithm.
%For example, finite difference techniques use truncated Taylor Series expansions to formulate their approximations.
%The overall accuracy and stability of a numerical simulation or calculation depends on both these kinds of errors.

Computational scientists have long been keenly aware of the importance of numerical
precision in designing algorithms \cite{hamming, goldberg}.
Because of this, essentially all fields utilizing modeling and simulation techniques on computers
have converged on community standards for representing data and making
performance versus accuracy trade-offs in, for example, time integration or numerical optimization.
While numerical methods may offer 3 or 4 digits of precision when run with
the IEEE single-precision (FP32) number format, scientists usually regard double-precision (FP64)
as the standard for scientific computing given its relative ease of achieving robustness and accuracy.
As the performance of a numerical algorithm using double-precision
has remained predictable at about half that of the same algorithm but using single-precision numbers,
community standards have not needed to be re-evaluated until a few years ago.

\begin{figure}[h]
    \centering
    %\begin{subfigure}{0.49\linewidth}
    %    \includegraphics[width=0.95\linewidth]{fig/flops_years_nvidia.pdf}
    %    \caption{NVIDIA}
    %\end{subfigure}
    %\begin{subfigure}{0.49\linewidth}
    %    \includegraphics[width=0.95\linewidth]{fig/flops_years_amd.pdf}
    %    \caption{AMD}
    %\end{subfigure}
    \includegraphics[width=0.7\linewidth]{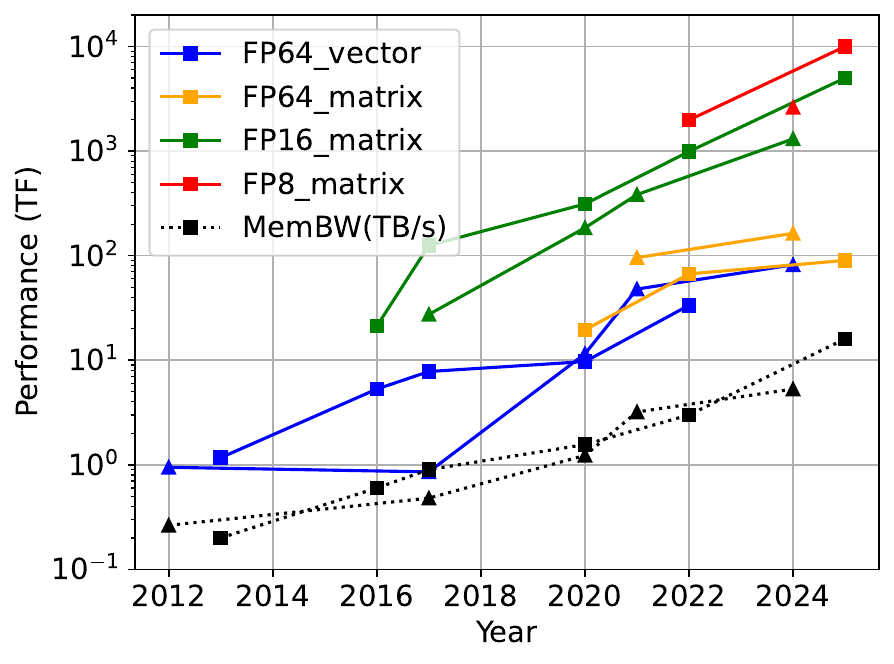}
    \caption{Floating point throughput and memory bandwidth of data-center GPUs over the years (note the logarithmic scale of the y-axis). Square markers represent NVIDIA GPUs while triangles represent AMD GPUs.}
    \label{fig:flops-years}
\end{figure}

Since the last several years, we observe two notable trends with regard to the performance of scientific applications. One comes from the evolution of hardware driven by artificial intelligence (AI) applications.
Increasingly, high-performance computing (HPC) hardware is moving towards more compute capacity for low-precision and mixed-precision computations, especially for matrix-matrix products (see Figure \ref{fig:flops-years}).
GPU vendors and other `AI hardware' companies are on track to disproportionately increase FP16, BF16, FP8 and INT8 performance, sometimes at the cost of FP64 performance.
The other is related to algorithmic advancements in computational science and engineering. The US Department of Energy's Exascale Computing Project and efforts in other countries have led to substantial advancement in mixed-precision algorithms and software, which enable the use of low precision arithmetic units in scientific computing.
Mixed precision calculation leads to performance gains and in many cases memory efficiency \cite{lehmann2022accuracy}. 
It can also lead to substantial improvements in energy efficiency \cite{sze2017efficient,sakamoto2020effectiveness}; for example, NVIDIA claims that a FP8 tensor core MMA FLOP requires only 0.06$\times$ the energy of a FP32 vector FLOP in matrix multiplication.
However, the use of reduced-precision formats (e.g., FP16 and BF16) can significantly affect solution accuracy \cite{sze2017efficient,lehmann2022accuracy} negatively.
Ltaief et al. in a recent study \cite{ltaief_responsibly_reckless_2022} refer to mixed precision algorithms as `responsibly reckless'. They point out that with carefully constructed algorithms, it is possible to gain savings in storage requirements and time in seismic imaging, environmental modeling and astronomical image reconstruction.

%\begin{figure}[h]
%    \centering
%    \includegraphics[width=0.85\linewidth]{fig/nvidia-mxp-energy-cropped.png}
%    \caption{Energy consumption of different precision formats on new NVIDIA GPUs (courtesy NVIDIA)}
%    \label{fig:formats_energy}
%\end{figure}

The evolution of double-precision (FP64) performance and half-precision (FP16) arithmetic performance over the last decade or so, from both the major GPU vendors (Figure \ref{fig:flops-years}), points to the widening gap between low-precision and high-precision throughput. Furthermore, newer GPUs from NVIDIA and Advanced Micro Devices (AMD) have even greater throughput for FP8, INT8 (8-bit integer) and even FP4 formats \cite{amd_tnp,nvidia_tmp_2024}.
%Specialized matrix-matrix multiplication units on modern GPUs are responsible for the disproportionate increase in low-precision (matrix) arithmetic preformance.
On legacy central processing units (CPUs), the maximum possible FP32 performance was about twice that of FP64 essentially irrespective of workload.
On modern hardware starting especially from the NVIDIA Volta generation, the introduction of tensor cores enables a disproportionately higher \emph{arithmetic compute} performance for lower precision formats such as INT8, FP16, BF16 and TF32 compared to FP32 and FP64.
Thus, if a compute-limited application can utilize lower precision to a significant degree, the maximum possible speedup is disproportionately high.
The evolution of floating point capabilities of hardware has been recently documented by Dongarra et al. \cite{dongarra_hardware_trends_2024}.
One risk is that actual scientific applications are frequently memory or network-bandwidth limited. In this case, the gain from the use of lower precision comes only from the ability to transfer more numbers in a given number of bits.
For example, if a memory bandwidth-limited application uses single precision (FP32) arithmetic instead of double (FP64), the maximum possible speedup is 2$\times$.

Table \ref{tab:precision} shows several popular precision formats and how their bits are allocated, their resulting ranges, and their machine epsilon $\epsilon$ value, i.e., the smallest number that can be represented such that $1 + \epsilon > 1$, where $\epsilon = 2^{-\text{mantissa}}$ \cite{cook_bfloat16_2018}. Brain Float 16, usually referred to as BFloat16 or BF16, was developed by Google specifically for deep learning training on their Tensor Processing Units (TPUs)  \cite{kalamkar2019study}, and has since become a widely adopted standard. BF16 has the same number of bits as FP32 to represent the exponent, but uses a reduced mantissa; therefore, it has the same numerical range as FP32, but can only represent 2-3 significant digits accurately as indicated by its epsilon value of 0.0078. 
TF32 was developed by NVIDIA for accelerating deep learning training on their Ampere architecture \cite{nvidia_tf32_2020}. It has the same range as FP32 but only uses 19 bits instead of 32 bits, making it computationally significantly less expensive. NVIDIA later introduced FP8 (with E4M3 and E5M2 variants) in support of training Transformer architectures on its Hopper architecture \cite{nvidia_fp8_primer_2023}. 

\begin{table*}
    \centering

    \renewcommand{\arraystretch}{1.5} 
    \caption{Summary of floating point precision formats (with BF16 example in the figure that follows)}
    \label{tab:precision}
    \begin{tabular}{lccccccl}
    \hline
    \textbf{Abbreviation} & \textbf{Total Bits} & \textbf{Sign} & \textbf{Exponent} & \textbf{Mantissa} & \textbf{Range} & \textbf{Epsilon} \\ \hline\hline
    FP64            & 64      & 1       & 11        & 52        & $\pm 10^{\pm 308}$  & $2.22 \times 10^{-16}$ \\
    FP32            & 32      & 1       & 8         & 23        & $\pm 10^{\pm 38}$   & $1.19 \times 10^{-7}$  \\
    TF32            & 19      & 1       & 8         & 10        & $\pm 10^{\pm 38}$   & 0.00097656 \\
    FP16            & 16      & 1       & 5         & 10        & $\pm 10^{\pm 5}$    & 0.00097656 \\
    BF16            & 16      & 1       & 8         & 7         & $\pm 10^{\pm 38}$   & 0.0078125 \\
    FP8 (E4M3)      & 8       & 1       & 4         & 3         & $\pm 448$           & 0.125  \\
    FP8 (E5M2)      & 8       & 1       & 5         & 2         & $\pm 57344$         & 0.25  \\ \hline
    \end{tabular}
    \vspace{5mm}
    \includegraphics[width=0.4\linewidth]{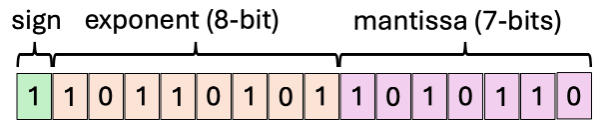}
\end{table*}

% expand on formats: mention TF32, etc.
%{\color{red} WB: Possible section on number representations and quantization (see Sec. VII.A p.22 on Reduced Precision \cite{sze2017efficient}).}

Computational scientists and engineers are concerned first with sufficient accuracy of simulation outputs, and secondly with the time it takes to compute these outputs.
Clearly, the accuracy of the outcome is the \emph{raison d'\^etre} of computer simulations. If the simulation accuracy is compromised beyond an acceptable level or the calculation fails due to faults like division by zero, there is no point of the simulation.
Secondly, in many domains, simulations are expensive, and computational resources limit the resolution and detail of the solution.
In these cases, scientists are very concerned about the resource efficiency and time-to-solution.
It is thus of great interest to estimate the impact of using mixed-precision algorithms in different domains of computational science on the accuracy and time-to-solution of simulations.
We leave the discussion on the energy-efficiency implications of mixed-precision algorithms to future work.

If the performance growth trends for low-precision and double-precision arithmetic continue to diverge (Fig. \ref{fig:flops-years}), it may eventually be possible, or even necessary, to directly approximate double-precision-accurate basic operations in software using fast hardware-accelerated low-precision arithmetic.
This is sometimes referred to by the community as `emulation' of high-precision arithmetic.

%\begin{wrapfigure}{r}{0.5\linewidth}
%\includegraphics[width=0.9\linewidth]{fig/flops_years_combined.pdf}
%\caption{Floating point throughput of NVIDIA and AMD data-center GPUs over the years}
%\label{fig:flops-years}
%\end{wrapfigure}

\begin{figure}[h!]
    \centering
    \includegraphics[width=0.6\linewidth]{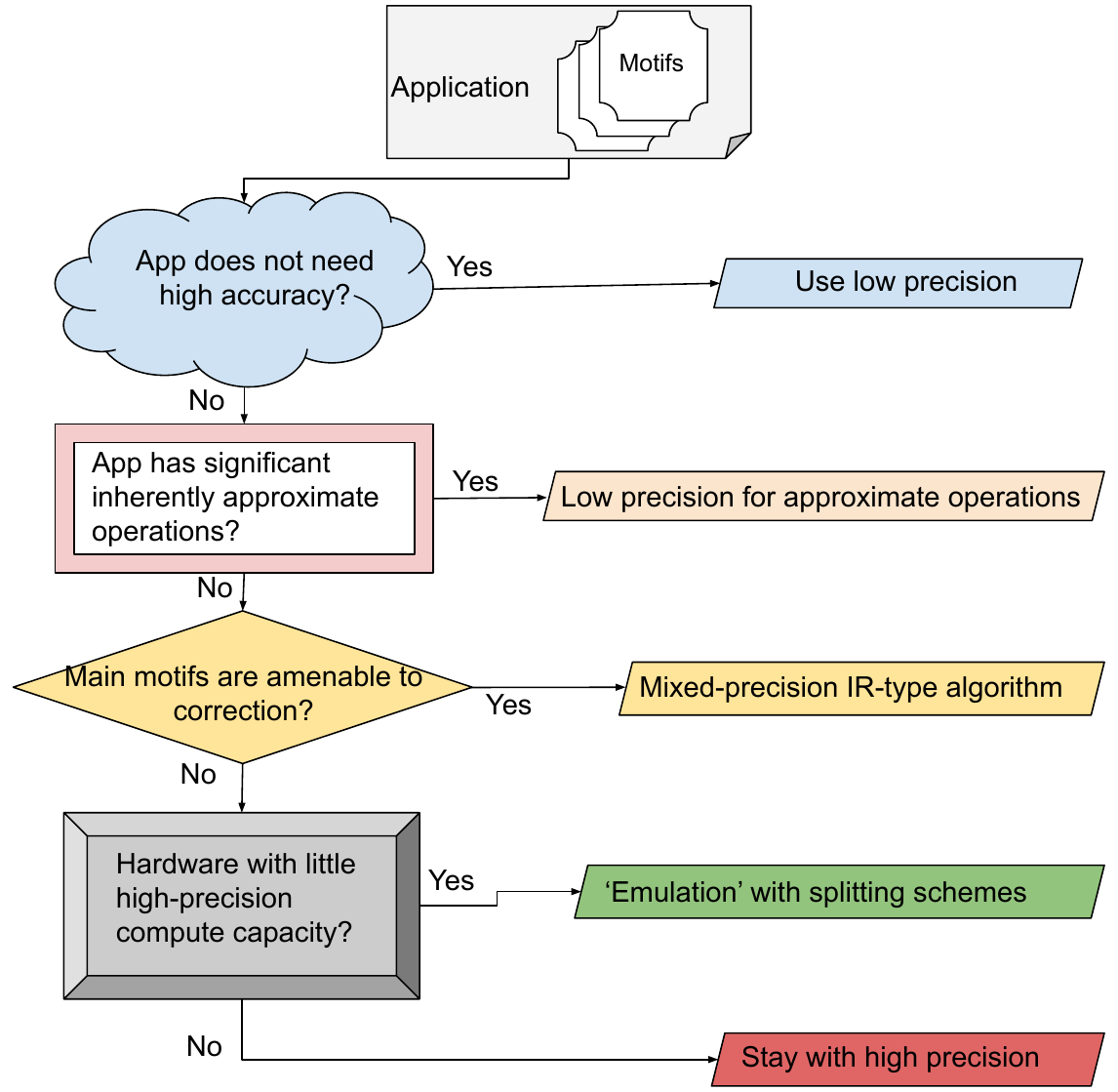}
    \caption{Considerations in using low or mixed-precision methods for scientific applications, revealing classes of mixed precision methods}
\end{figure}

All this should prompt computational scientists and engineers to investigate whether low-precision or mixed-precision calculations can be used for their research to improve time-to-solution and energy efficiency.
After a review of applications, common numerical computational motifs, algorithms, and software libraries, the following are our observations.

%\begin{description}
\begin{enumerate}
\item \emph{\textbf{Impact.}}
%\subsubsection{Impact}
If mixed-precision techniques can be effectively utilized for a significant portion of scientific computing, it will have impact on the pace of scientific progress and the resources needed to achieve it.
The use of mixed-precision will impact time-to-solution, memory footprint and energy efficiency of an application. Massive gains in time-to-solution and energy efficiency are likely in some domains, with some gains possible in most domains.
This is especially true on new and upcoming hardware platforms as of writing.

\item \emph{\textbf{Domain-specificity.}}
%\subsubsection{Domain-specificity}
Whether and how mixed precision is used in an application depends on the scientific domain and the problem.
For some domains in some parts of the application, due to inherent input uncertainties, a lower precision can be utilized without affecting the usefulness of the solution.
Often, low precision can be used only in certain parts of the algorithm or certain regions of the spatio-temporal domain under study without affecting the overall solution.
Sometimes, simple variable or equation transformations can help retain accuracy while using lower precision formats.
In other areas, the solution obtained from a lower-precision calculation needs to be corrected in some way, typically through some kind of iterative refinement.

\item \emph{\textbf{Identifiable compute motifs as the driver.}}
%\subsubsection{Identifiable compute motifs as the driver}
Identification of computational motifs can help adopt the right mixed precision algorithms, and in some cases good implementations are already available.
Over the past decade or so, there has been research on mixed-precision algorithms for many computational motifs including dense and sparse factorizations, iterative solvers, preconditioners, multigrid methods, least-squares problems, Fourier transforms and nonlinear solution methods.
New algorithms have been tested, theories of accuracy derived, real-world accuracy measured, and speedups estimated or measured. Software infrastructure has been developed (often by investments from government agencies like the US Department of Energy) to enable computational scientists and engineers to explore the applicability of mixed-precision algorithms in their domains.
It is also possible to emulate high-precision arithmetic using low-precision hardware, especially for dense matrix-matrix multiplication (GEMM).

\item \emph{\textbf{Mixed precision can go beyond just arithmetic compute throughput.}}
%\subsubsection{Mixed precision can go beyond just arithmetic compute throughput}
What remains is for applications and problems to be matched with existing research on mixed-precision algorithms and software.
In some cases, there may be juicy low-hanging fruit waiting to be picked - the acceleration of floating-point compute-intensive applications that already run on large supercomputers that do not currently use mixed-precision methods. Such applications can leverage the massive throughput of `tensor' or `matrix' processing units on modern GPUs and other `AI hardware'.
Further, there are a host of other applications that use floating-point computations, but are not limited by compute throughput but rather by memory or communication bandwidth. These applications are waiting for the right algorithms and software (and perhaps, eventually, hardware) technology to leverage low-precision memory and communication operations to improve time-to-solution. Some of the algorithms and software required for this already exist.

\item \emph{\textbf{New crop of algorithmic innovation.}}
%\subsubsection{New crop of algorithmic innovation}
Finally, there are ongoing efforts to utilize machine learning and AI methods for accelerating simulations. Deep learning frameworks typically automatically take advantage of low-precision hardware, though newer architectures such as neural operators may need some algorithm and code changes.
Further, AI hardware can be utilized by some basic computational motifs, such as matrix-matrix multiplication, for high-precision computations by `emulating' these on the low-precision hardware units.
\end{enumerate}
%\end{description}

In view of these five observations, we recommend a co-design approach involving application scientists, math library developers and computer scientists to incorporate effective mixed-precision strategies into widely-used scientific applications, thereby accelerating scientific progress at greater energy efficiency.

In this survey, we undertake a review of
\begin{itemize}
    \item scientific applications that currently take advantage of mixed-precision algorithms or projects that have studied that impact of using mixed-precision algorithms in their programs,
    \item methodologies and algorithms that have been developed to enable mixed-precision implementations of commonly-used computational motifs,
    \item the state of the software ecosystem in terms of availability of optimized mixed-precision algorithms,
    \item some recent techniques to `emulate' certain double-precision arithmetic operations using low-precision formats, and
    \item briefly, the impact of mixed-precision methods on resource utilization.
\end{itemize}
Some previous papers and reports have addressed a few of the above points, primarily numerical linear algebra and associated libraries, including by the United States' Exascale Computing Project \cite{ecp_mxp_advances_2021,abdelfattah_survey_2021,ecp_mxp_kit_2024} and Higham and Mary \cite{higham_mary_2022}.
However, as of writing, there is no work that addresses all of the above points in the literature.

In most of this work, we focus on science applications that primarily utilize domain knowledge and numerical methods, as these are the workhorses of computational science and their inner workings are relatively well-understood by experts. It is largely in this context that the potential advantages of mixed-precision approaches are unexplored.
However, the use of low and mixed-precision methods is common in deep learning and artificial intelligence (AI), and the use of these methods in computational science is one way to take advantage of low-precision hardware.
Thus, we briefly discuss the use of AI methods to utilize low-precision arithmetic in scientific research.

Finally, we return to our perspective on mixed-precision algorithms for science and our recommendations on future research directions in this regard.
Eventually, our goal is to make the best use of currently-available and upcoming hardware platforms to deliver meaningful and impactful outcomes for computational science.

\section{Application areas}
\label{sec:appls}

In this section, we review some application areas relevant to the Department of Energy and explore to what extent and how they use mixed-precision techniques.
Our choice of application domains and projects is somewhat subjective. We choose them based on our own expertise, the research programs at the largest supercomputing centers, and also those applications that are widely in use and depend on compute-limited numerical operations.

As mentioned in the introduction, many scientific and engineering applications tend to be limited by memory bandwidth, thus limiting the possible speedup that can be gained by the use of lower precision formats. 
In this context, we are interested in identifying those scientific applications that are limited by compute throughput rather than memory or communication bandwidth or latency.

At the Oak Ridge Leadership Computing Facility (OLCF), the Center for Accelerated Application Readiness (CAAR) was created to coordinate the substantial effort it takes to bring existing scientific applications to the point of effective use of heterogeneous CPU-GPU supercomputers.
We include the scientific applications under the Frontier CAAR for their use of mixed-precision approaches in our survey. These were the applications identified to be optimized for exascale performance during the development of the supercomputer Frontier.\cite{budiardja_caar_2023}
At the end, we identify cross-cutting opportunities and challenges that emerge across application domains.

We can already identify one broad mathematical tool that unites most of the application domains in scientific computing: nonlinear partial differential equations (PDEs) are commonly used in the physical sciences to model phenomena ranging from fluid dynamics and solid mechanics to electromagnetism and quantum physics.
We discuss in section \ref{motifs:nonlinear} some of the considerations in using lower precision formats in nonlinear solvers.
In addition, certain physical phenomena are `modelled' rather than `resolved'. This means that they are accounted for in the overall simulation using simplified models rather than first-principles rigorous treatment.
Examples include `averaged' or `sub-grid scale' turbulence models in aerodynamics and hydrodynamics, and `parameterizations' of variables such as humidity in certain weather models.
Such simplified models are also targets of optimization using lower precision.

% \subsection{Aerodynamics}
\subsection{Computational Fluid Dynamics}

Computational Fluid Dynamics (CFD) encompasses a wide range of methods and types of flows, from Reynolds-Averaged Navier-Stokes (RANS) methods, Large Eddy Simulation (LES), Direct Numerical Simulation (DNS) methods, as well as Lattice Boltzmann methods, among others. Such methods are used to model a wide variety of flow-fields, from hydrodynamics and aerodynamics to multi-phase flows in nuclear power, astrophysical flows, etc. 

Brogi et al. \cite{brogi2024floating} perform a thorough investigation of the accuracy and convergence of reduced precision CFD simulations using OpenFOAM, a widely used open-source CFD code, which is primarily based on the finite volume method. They study four benchmark cases: lid-driven cavity flow; isotropic decaying turbulence; the one-dimensional gas shock tube problem; a starting compressible jet; and finally a realistic simulation of a volcanic plume.
They test each of their models using double precision, single precision, and mixed precision -- where all the code is compiled in single precision except for the linear algebra solvers, which use double precision. Their results show that single precision is sufficient for most laminar flows, but not always for turbulent flows -- in which case the reduced precision compromises the performance of the linear algebra solvers. In regards to performance gains by reduced precision, they show speedups for single precision studies ranging from 1.29$\sim$1.87x with respect to double precision for the lid-driven cavity flow case. For the starting jet benchmark, they investigate mixed precision for different types of linear algebra solvers. The optimal scenario uses mixed precision with a Block Jacobi solver and yielded a speedup of 2.43x with respect to double precision. 

Lattice Boltzmann methods (LBM) have emerged as a fundamentally different approach to simulating fluid flow, based on the concepts of discrete particle dynamics and cellular automata \cite{frisch1986lattice}. Lehman et al. \cite{lehmann2022accuracy} study Lattice Boltzmann methods at 64-bit, 32-bit, and 16-bit precision for six increasingly complex test cases: Poiseuille flow, Taylor-Green vortices, Karman vortex streets, lid-driven cavity flow, immersed-boundary micro-capsule in shear flow, and modeling the impact of a raindrop using a volume-of-fluid approach. Their approach generally uses FP16 for storing fluid populations, while performing the arithmetic operations using FP32. By analyzing the range of numbers used for LBM, they realized they could develop a custom FP16, called FP16C, which increases the mantissa by one bit for improved accuracy. They generally show that there is a negligible difference in accuracy between FP32 and FP64, and are able to achieve the theoretical speedup of 1.8x by using mixed FP32/FP16C.

On the other hand, LBPM (Lattice Boltzmann Methods for Porous Media) \cite{lbpm_2021} is a CAAR application that uses Lattice Boltzmann methods to model flow through porous media and other related transport processes \cite{budiardja_caar_2023}.
There is no use of mixed-precision algorithms in the code.

Walden et al. describe \cite{walden_unstructured_mixed_2019} mixed-precision kernels on the state-of-the-art finite volume unstructured grid aerodynamics code FUN3D running on the Summit supercomputer at OLCF. They document large-scale Reynolds-averaged turbulent flow simulations with over a billion grid points using two-precision and three-precision sparse linear iterations.
The linear solver they use is the multicolor block Gauss-Seidel iteration.
In their double-single linear solver implementation, the right-hand side vector $-\bld{r}$ and the diagonal blocks of the Jacobian matrix $\frac{\partial\bld{r}}{\partial\bld{u}}$ are held in double precision, while the solution vector $\Delta \bld{u}$ and the off-diagonal blocks of the Jacobian matrix are held in single-precision.
In their double-single-half implementation, the off-diagonal blocks of the Jacobian matrix are instead stored in half-precision. Since the maximum possible FP16 value is 65504, they scale the off-diagonal values by the ratio of 65504 and the maximum off-diagonal value.
They found challenges in assembling the off-diagonal blocks directly in low precision. Therefore, since making a copy is wasteful especially in terms of memory footprint, a sophisticated kernel is needed to convert the off-diagonal part of the matrix to low precision in-place.
They find that the double-single-half linear solver's convergence stalls at around 35 iterations for a representative matrix from a steady-state wing-body simulation on a single GPU. However, since the typical number of iterations actually performed in a nonlinear iteration is around 15, this does not impact the overall solver convergence.
A 1.2$\times$ speedup is observed for the full simulation going from the double-single implementation to the double-single-half implementation. If we assume a further 1.2$\times$ speedup going from a fully double-precision implementation to their double-single implementation, the overall speedup over double would be about 1.44$\times$.
For a time-dependent detached eddy simulation on 552 GPUs, a 1.11$\times$ speedup is observed since the linear solver constitutes a smaller proportion of total time for this type of simulation.

As early as 2001, Gropp et al. \cite{petsc_fun3d_2001} reported on FUN3D with a PETSc (see \ref{sec:other_libraries}) backend to use matrix-free Newton-Krylov solvers with scalable Schwarz preconditioners from PETSc.
They experimented with storing and accessing the preconditioner using single precision, and observed a nearly 2$\times$ speedup for the linear solver phase (not the entire application). However, they do not discuss this any further beyond an experiment to prove that the linear solver performance is limited by memory bandwidth.

Cholla \cite{cholla_2015} is a compressible flow (CFD) code for use in astrophysics problems and a Frontier CAAR application. It uses a finite volume method and explicit time stepping, second or third order accurate in space and second order accurate in time, to solve the partial differential equations (PDEs) of fluid dynamics.
The performance of this is kind of code is limited by memory bandwidth \cite{budiardja_caar_2023}, similar to FUN3D \cite{petsc_fun3d_2001}.
Since it only uses Cartesian grids, it has very regular memory access. Thus, its performance likely closely follows the memory bandwidth limit even more so than FUN3D, which uses unstructured grids.
As of writing, there are no known efforts to utilize mixed-precision numerics to accelerate Cholla.
Unlike implicit temporal discretizations which typically have a nonlinear and linear solve steps within each ordinary differential equation (ODE) stage, it is more difficult to use lower precision in explicit time stepping and still retain high accuracy.

While we are on the subject of medium order accuracy methods in CFD that use explicit time stepping, an interesting approach to mixed precision has been adopted by Field et al. \cite{field_weno_2021}.
They use a Weighted Essentially Non-Oscillatory (WENO) method to compute accurate solutions to a linear hyperbolic PDE in astrophysics and advance the solution using a strong stability-preserving Runge-Kutta (SSPRK) explicit method.
They compute the WENO weights in low precision, and as long as the weights add up to 1.0, this should not compromise accuracy (though one still needs to be careful in regions with strong shocks).
In their experiments, they retain sufficient accuracy in the region of interest, though there are errors introduced in some unimportant regions. They obtain a speedup of 3.3$\times$ by switching the weight computations from quad precision to double precision on an NVIDIA V100 cluster. (The original code uses all quad computations to preserve long-time stability in time integration.)
We document some progress in mixed-precision ODE solution methods in section \ref{sec:ode}.

The GESTS (GPUs for Extreme Scale Turbulence Simulations) project was chosen for the Frontier CAAR and consists of a direct numerical simulation (DNS) code for simulating turbulent flows with very high resolution by accurately solving the PDEs of fluid flow, the Navier-Stokes equations.
Its primary computation is 3D fast Fourier transform (FFT) which is used in the pseudo-spectral discretization of the PDEs.
Its performance is limited by communication in 3D FFTs \cite{yeung_2019}. The MPI communications take approximately 70\% of the runtime, while local FFT computations take 20\% \cite{budiardja_caar_2023}.
Again, as of writing, there have been no known attempts to leverage mixed-precision methods to accelerate this application.

Karp et al. present very interesting recent investigation on the accuracy impact of using mixed precision methods in DNS and large eddy simulation (LES) of turbulent and transition flows \cite{karp_mxp_scale_resolving_2025}.
They present four different test cases, including three incompressible and one compressible, and four different CFD codes with diverse numerical methods.
They found that single precision can, by and large, be used for turbulent flow simulations without impacting the important characteristics of the solution, with some caveats.
First, for DNS, the FP32 solution sometimes shows spurious oscillations for fluctuating terms, especially in higher order moments. This can typically be fixed by carrying out one time step in FP64 at the end of the simulation before postprocessing the results.
We remark that this is applicable for spatial averaging and other ensemble averaging; the authors do not report on how this applies to temporal averaging.
Second, they note that the spectral code is more sensitive to precision than the finite element based codes. 
Third, they observe (eg. in LES of flow over a cylinder) that FP32 simulations are, in general, physically realistic but the drag coefficient values differ starting with the third significant digit.
Fourth, time-averaging over smaller windows of time (eg., 100 time units) tend to show more difference between FP32 and FP64 runs.
Finally, for the compressible flow case, they did not consider a fully FP32 solver, but one where the the state is rounded to FP32 before computing the fluxes, and the fluxes are again rounded to FP32 before stepping in time.
They give the following recommendations:
\begin{itemize}
    \item For simulations with small time steps, the times and associated weights etc. should be stored in FP64, since adding small time steps sometimes leads to issues with FP32. This also impacts running time averages of quantities of interest.
    \item As noted by other researchers, accumulations, eg. in dot products, must usually be performed in FP64.
    \item In high order discretizations, geometric mapping terms may need to be stored in double precision.
\end{itemize}
Karp et al. found that simulations that use FP16 are more prone to spurious solutions. With some codes, using FP16 only for the convective term works well. But when the code is changed or the state variables are also represented in FP16, the errors grow too large.
As one would expect, these issues are also observed with the FP8 formats E5M2 and E4M3.

Wilfong et al. \cite{wilfong_igr_2025} ran a compressible turbulent flow simulation the extremely large scale of 100 trillion grid points across all of the exascale Frontier machine. They used a high-order finite volume discretization on structured grids, with a novel shock stabilization scheme, and explicit Runga-Kutta time stepping.
In one such simulation, they showed that there was no qualitative difference in FP64 and FP32 solutions of exhaust plumes of spacecraft engines.
FP16 simulations, however, showed larger plumes and backward heat flux regions, and also some non-physical artifacts. The authors believe this is due to accumulation in FP16.

\subsection{Weather and climate simulations}
\label{sec:appl:weather}

Saffin et al. \cite{saffin_reduced-precision_2020} studied reduced-precision `parameterizations', or simplified models, and how well the resulting rounding errors are masked by inherent model uncertainty. They mention that `the majority' of ECMWF's Integrated Forecasting System (IFS) was expected to be transitioned from double to single precision for a 40\% reduction in cost and no noticeable change in forecast skill. (Indeed, Lang et al. \cite{lang_ecmwf_fp32_2021} reported that the use of single precision in the IFS enabled an increase in vertical resolution, and thus actually \emph{increasing} the forecasting skill.)
In their work, they solve the hydrostatic equations in vorticity-divergence form, and specific additional physics such as humidity are modelled by simplified models called parameterizations.
They study the impact of using low precision using emulation and thus don't speak to time-to-solution improvements, only the effect on accuracy. This allows them to test the simulation behaviour for a wide range of precision with the number of mantissa bits ranging from 1 to 52 for a fixed number (11) of exponent bits.
They mention in passing that using 8 exponent bits (as in FP32) is sufficient, but the code crashes if this is reduced to 5 (as in FP16).
A methodology known as `stochastic perturbation of parameterization tendencies' (SPPT) is used to simulate model uncertainties.
An important concept in this work is that a lower precision is acceptable if the \emph{probability distribution of an ensemble of runs} in that precision is indistinguishable from an ensemble of runs in double precision.
In this case, the runs in the ensemble use different realizations of the perturbation of the parameterization tendencies. 
Note, however, that the specific SPPT used here significantly overestimates the model uncertainty compared to the reference implementation used on the IFS.

Saffin et al. \cite{saffin_reduced-precision_2020} find that using 10 or more mantissa bits, the similarity between the probability distributions of the low precision ensemble and double precision ensembles is acceptable, roughly between 81\% and 83\% similar for lead times from 1 to 30 days.
Using 8 mantissa bits, the similarity coefficient drops to 76\% at a lead time of 10 days.
There is another interesting finding:
For intermediate precision from 27 to 51 mantissa bits, the difference with respect to deterministic double precision geopotential height results can go from insignificant to several meters suddenly at a particular forecast lead time (eg., about 13 days for 51 bits).
The authors posit that this is because of the convection parameterization that switches on and off based on values at previous time steps.
The authors propose modifications of two of the variables - moist static energy in the convection parameterization and temperature in the surfaces-fluxes - that significantly improve the low-precision (8 mantissa bits) results.
These modifications depend heavily on the specific physical considerations of the parameterizations and physically-relevant ranges of values.
For example, expressing temperature in Celsius instead of Kelvin for computing the surface-fluxes parameterization improves accuracy due to additive movement of the range of values to a lower absolute number.
Finally, the authors conclude that it is possible to use half-precision arithmetic for all the parameterizations and still achieve predictions that are statistically very similar to double-precision results.

It has been argued by Ackmann et al. \cite{appl_wc_ackmann} that mixed precision can be used for solving the shallow water equations for weather simulations. Specifically, they hypothesise that reduced precision can play a role in the preconditioner of the linear solver for the elliptic equation in a semi-implicit time discretization.
They use a preconditioned restarted Generalized Conjugate Residual (GCR) solver, which is mathematically equivalent to preconditioned flexible GMRES \cite{saad_iterative_methods}, with a subspace size of 3.
%It may be noted that the relative residual tolerance for the linear solver is $10^{-5}$.
The preconditioner is itself a preconditioned Richardson iteration with a custom `physics-informed' preconditioner. Each Richardson iteration requires the inversion of a tridiagonal matrix.
The mixed-precision solver is tested on the Rossby-Haurwitz wave problem with wavenumber 4 (RHW4) and a zonal flow past the Earth's orography. Differences from reference solutions are studied.
%One thing to note is that their `half precision' format is emulated and uses 10 significand bits and 11 exponent bits, for a total of 22 bits.
%The mantissa is thus the same length as FP16, but the exponent is the same length as FP64.
Salient observations from Ackmann et al. \cite{appl_wc_ackmann} include:
\begin{enumerate}
    \item Solver convergence stalls if the entire shallow-water model is reduced to single-precision. However, the obtained model solutions are still physical, though different from the double-precision solutions. The difference is comparable to that arising from a small random perturbation to the initial conditions.
    \item When the iterative updates in the unknown solution (fluid thickness $\Phi$) become comparable to machine epsilon, the solver stalls. Using a double-precision variable for accumulating the increments restores most of the convergence.
    \item In the mixed-precision implementation, various variables are held in double precision and some operations are performed in double precision depending on experiments on the accuracy of the overall simulation.
    %For example, the solution variable in the GCR solver is held in double precision, and updates to it are up-cast, as in $\Phi_{\nu+1} := \Phi_\nu + \beta p_\nu$.
    \item A lot of the accuracy and stability consideration is driven by the ill-conditioning of the Helmholtz operator at the poles of the earth. They observe that precision reduction close to the poles causes spurious behaviour.
\end{enumerate}
At the nonlinear solver level, the residual ($\bld{r}$ in equation \eqref{eq:newton}) is computed in double precision but then held in single precision. The Jacobian matrix of the momentum equations is computed in single precision. As for the elliptic operator matrix for the mass conservation, they split the solution variable and operator matrix into two single-precision variables
\begin{align}
    \hat{\Phi} &:= \text{trunc}(\Phi^0) \\
    \tilde{\Phi} &:= \text{trunc}(\Phi^0 - \hat{\Phi})
\end{align}
where $\Phi^0$ is the initial value of the solution vector in double precision, and correspondingly compute the matrix in two parts using the spatial derivatives of $\hat{\Phi}$ and $\tilde{\Phi}$. This split into two single-precision vectors is related to multiword arithmetic \cite{fasi_multiword_2023} and is done only away from the poles; at the poles, full double precision operations are used.
This method yields a sufficiently accurate simulation for several days of lead time, especially away from the poles.
This single-double mixed precision solver attains a 30\% speedup over the double-precision variant. For larger problem sizes, they make a second-hand claim of speedups approaching 40\%.

In addition, they test a version with their half-precision format used for the preconditioner application away from the poles. Though there is a stall in the last linear solver iteration in all time steps, the solution retains the same qualitative features as the double-precision solution and difference magnitudes are not very large. For this half-single-double mixed-precision solver, they \emph{estimate} a 3.3$\times$ speedup for the elliptic part of the code. These speedups are consistent with a fully memory-bandwidth limited performance regime.

Kl\"ower et al. \cite{kloewer_shallowwater_fp16_2022} show that 16-bit calculations are sufficient for earth-system models solving the shallow water equations.
They obtain a 3.6$\times$ speedup on the Fujitsu A64FX processor for their memory bandwidth limited workload.
This is achieved with the help of innovative precision analysis procedures, and the method of compensated summation.
\begin{itemize}
\item An analysis of the range of numbers encountered in a simulation is performed by replacing the scalar type (eg. FP32) with `Sherlog', an overloaded numeric type that logs the the result of every calculation performed in a histogram.
Non-dimensionalizing the spatial gradients with the grid spacing helped reduce the occurence of numbers outside the FP16 range.
\item Another overloaded numeric type records a stacktrace whenever the result of an arithmetic operation satisfies a pre-specified condition such as \texttt{f(x) = 0 < abs(x) < floatmin(Float16)} (Julia code).
This revealed that the addition of tendencies to the solution variables (particularly, velocities) was producing subnormal numbers that are poorly handled in FP16 on several hardware platforms.
To address this, the velocity variables are scaled with a constant $s=10^6$, which reduces subnormals substantially while avoiding overflows.
\end{itemize}
However, it is unclear how well these methods generalize to, for example, other regimes of simulation parameters, non-uniform grids, and more realistic governing equations.

They also use the technique of compensated summation to reduce rounding error accumulation from time integration in low precision. This is an alternative to using higher precision to accumulate the sum, and may achieve a higher speedup.
\begin{greenbox}
\textbf{Compensated summation}
\begin{algorithmic}[1]
    \Require Finite sequence $(a)_{i=1}^n$ to be summed
    \State $c \gets 0$ \Comment{Compensation variable}
    \State $s \gets 0$ \Comment{Sum}
    \For {$i < n$}
        \State $y \gets a_i - c$
        \State $t \gets s + y$
        \State $c \gets (t - s) - y$
        \State $s \gets t$
    \EndFor
    \State \Return $s$
\end{algorithmic}
Using this method, Higham \cite{higham_summation_1993} states, attributing to Knuth, that the backward errors $\mu_i$ are bounded as
\begin{equation}
    |\mu_i| \leq 2u + \mathcal{O}(nu^2),
\end{equation}
where $u$ is the low-precision unit roundoff and the $\mu_i$ are defined by ($\hat{s}$ is the computed sum)
\begin{equation}
    \hat{s} = \sum_{i=1}^n (1+\mu_i)a_i.
\end{equation}
\end{greenbox}
With compensated summation, Kl\"ower et al. \cite{kloewer_shallowwater_fp16_2022} achieve a 3.6$\times$ speedup on large grids, compared to $2.75\times$ with mixed precision summation (accumulation in FP32).

\subsection{Quantum chemistry}
\label{area:quantum}

Relatively early work includes the electronic structure program TeraChem, in which calculations were performed in mixed double and single precision \cite{titov_quantum_2013}.
Depending on certain domain-specific bounds, certain electron repulsion integrals were computed in double precision while others in single. On old NVIDIA GPUs from the Fermi and Kepler generations, they claim a nearly 2$\times$ speedup from this approach while retaining nearly the same accuracy as a full double-precision calculation.
Note that on the Kepler K20, the single-precision throughput was three times the double-precision throughput.

Das et al. contributed a finite element methodology to solve the Kohn-Sham equations of density functional theory (DFT) with a code called DFT-FE. This code achieved 33\% of the benchmark performance on the Summit supercomputer \cite{das_dftfe_2019} and used mixed-precision techniques.
The code solves the Kohn-Sham equations in a finite-element polynomial basis by iterative solution
of an electron orbital eigenvalue problem in a Chebyshev filtered subspace of the full basis.
Careful consideration of the precision used during each step of applying the linear (Hamiltonian) operator,
and in solving the eigenvalue problem
%The adoption of mixed-precision strategies for dense matrix calculations in the basis vector orthonormalization and Rayleigh-Ritz procedures results in about
results in about 85\% of floating-point operations able to be done in FP32 with the remaining in FP64.
Since the contribution of FP32 computations to the electron density tend to zero as the nonlinear eigenvalue solver iteration converges, the ground-state solution retains FP64 accuracy.
In its Chebyshev filtering phase, this application uses low precision (FP32) for all-to-all communications via \texttt{MPI\_Allreduce}, reducing communication cost by a factor of 2, while ground-state solutions are reported to retain FP64 accuracy.
There are two strategies used:
\begin{itemize}
    \item Compute local GEMM-like operations on each GPU in double precision, while communicating neighborhood information and computing the off-diagonal contribution in single precision.
    \item In the Rayleigh-Ritz eigenvalue procedure, compute the (projected) Hamiltonian matrices in double precision only for fractionally-occupied or unoccupied eigenstates, while performing the projection in FP32 for fully-occupied eigenstates.
\end{itemize}
Due to these mixed-precision schemes, they attain a 2$\times$ speedup over fully double-precision computations in four of the seven computational steps in each nonlinear iteration. These are the steps that depend on dense GEMM operations on each GPU, and globally scale as $\mathcal{O}(MN^2)$, where $M$ is the number of finite element degrees of freedom while $N$ is the number of electron orbitals of the chemical system studied.
In linear algebra terms, $N$ is the number of independent right-hand sides in the system of equations.
The steps that depend on matrix factorization are performed in double precision. However, these only depend on $N$ and scale as $\mathcal{O}(N^3)$. These are done on the CPU and are claimed to not be the dominant cost - globally, $N \equiv 60,000$ while $M$ can be up to hundreds of millions of degrees of freedom.

From memory bandwidth considerations, the overall speedup would be just under 2$\times$.
However, the presence of many right-hand sides (in the thousands) makes the problem compute-bound,
resulting in a speedup higher than this memory bandwidth-limited estimate.
On OLCF Frontier, the code achieved 43.1\% of the peak theoretical FP64 FLOP/s \cite{das_dftfe_2023}.

Dawson et al. \cite{dawson_quantum_mxp_2024} have applied the Ozaki I scheme (section \ref{sec:ozaki_1}) to emulate higher precision computations in quantum chemistry.
They use a density matrix purification method to calculate the single-particle density matrix from a given mean-field Hamiltonian. This method is similar to the idea of Chebyshev polynomial iteration used in DFT-FE. 
The difference is that they apply the Hamiltonian $\bld{H}$ to the density matrix $\bld{K} $(\cite{dawson_quantum_mxp_2024} Eq. 5), where DFT-FE applies $\bld{H}$ to the wavefunction coefficients $\bld{C}$. $\bld{K}$ has the same dimensions as $\bld{CC}^T$, and thus applying $\bld{H}$ to it involves much more work.

They make the case \cite{dawson_quantum_mxp_2024} that fully double precision calculations are unnecessary for reasonable convergence criteria of the Self-Consistent Field (SCF) iterations, while single precision is insufficient.
When using as low as 24 mantissa bits for accumulation, they recommend carrying out an additional iteration (McWeeny step) in double precision to restore accuracy.
They give example errors for specific calculations, but not a general theory - so there is a need for a better theory of approximation error in these iterative schemes.
Next, from their results of applying the Ozaki I scheme, we note that the BF16 format cannot provide enough accuracy using a reasonable number of splits, while pure FP16 does better. FP16 TC (FP16 multiplication and FP32 accumulation on NVIDIA tensor cores) provide sufficient accuracy using just 2-4 splits.
Thus, the potential for speedup is very good. Unfortunately, however, the authors do not report any measurements of speedup or energy savings.
The findings of this work reinforce our take-aways from DFT-FE: domain-specific knowledge of sparsity and solvers that apply iterative schemes are especially good targets for mixed precision optimizations.

LSMS \cite{lsms_gpu}, or Locally Self-consistent Multiple Scattering, is a Frontier CAAR application that solves the Schr\"odinger equation of electrons using density functional theory (DFT).
In its main production mode, its runs are dominated by dense double-precision complex matrix operations, particularly LU factorization and thus matrix-matrix products \cite{experiences_exascale}.
Thus, a significant portion of this code's runs may be floating point compute-limited, and this makes it a candidate for investigating mixed-precision algorithms.
However, so far, there are no published attempts at utilizing mixed-precision here. Attempting to use mixed-precision LU factorization with iterative refinement (see subsection \ref{sec:dense_factorization}) may be a low-hanging fruit for such workloads.
There have been preliminary studies on using single precision dense complex linear solves in LSMS, and preliminary results indicate that the results have acceptable accuracy [private communication].

Tian et al. report \cite{tian_dmrg_2022} a mixed precision implementation of a two-site Density Matrix Renormalization Group (DMRG) calculation.
They start the iterative simulation with single precision iterations, and finish with a few double-single mixed precision iterations.
The latter use double precision computation for some operations to retain accuracy.
On Intel Xeon Gold 6126 CPUs, their mixed precision simulation achieves a speedup of 1.29$\times$ to $1.81\times$ depending on the chemical species being studied, while having an error of less than 0.013\% in the system energy relative to the fully double precision solution.

The General Atomic and Molecular Electronic Structure System (GAMESS) project is a quantum chemistry package written primarily in Fortran with newer development in C++. This application was among in the Frontier CAAR projects. The main computational operation in GAMESS is dense linear algebra, particularly symmetric eigenvalue solver \cite{experiences_exascale}.

NuCCOR is a quantum physics application that carries out ab-initio simulation of atomic nuclei. It solves the time-independent Schr\"odinger equation using the coupled cluster method \cite{budiardja_caar_2023}.
It is written in Fortran 2018, and was also one of the Frontier CAAR applications.
The coupled-cluster method yields an eigenvalue problem, which NuCCOR solves using iterative Krylov-subspace methods such as Arnoldi and nonsymmetric Lanczos. The smallest eigen-pairs are extracted to a specified precision.
Distributed block-sparse tensor contractions dominate the computational cost. This leads to dense tensor contractions as the local operations on each MPI rank.
As of writing, NuCCOR does not use mixed-precision algorithms.

A key ingredient in applying mixed-precision arithmetic seems to be identifying a problem
structure where the converged solution has a subset of components that tend to zero.
Standard dense methods do not typically incorporate this kind of knowledge.  However, these examples show that this kind of structure can appear in iterative methods that converge on
active subspaces.

\subsection{Computational genomics}

CoMet is a software tool for vector similarity comparison. It employs two measurement types: CCC (Customer Correlation Coefficient) and Proportional Similarity metric \cite{Joubert_gordonbell,Joubert_ccc, Joubert_pf}. This software finds application primarily in the biological domain. 

Genome-Wide Association Studies (GWAS) and Quantitative Trait Loci (QTL) analyses aim to identify genetic variants that contribute to individual phenotypes, including susceptibility or resistance to diseases. However, studies have shown that the risks for many complex human diseases stem from non-additive, epistatic interactions between multiple genes. These interactions involve combinations of genomic variants that collectively lead to disease, rather than single genetic factors acting independently.

%{\color{red} AK: What is an epistatic interaction? Can we either define the term or leave it out? epistatic mean it involves two or more genes that could be none-additive or greater than additive. Like if some gene make me have big nose, and another also make me have big nose, I may not have bigger nose (none-additive), vice versa }

Each Single Nucleotide Polymorphism (SNP) represents a difference in a single DNA building block, called a nucleotide. For example, an SNP may replace the nucleotide cytosine (C) with thymine (T) in a certain stretch of DNA. Brute force search through all possible epistatic interactions is infeasible. Hence, a method of building an SNP network using correlation measurements and clustering the network for smaller interaction sets has been developed to significantly reduce the search space.

The CCC measurement is used to find correlations between each SNP given a set of individuals exhibiting a particular phenotype, effectively measuring whether these SNPs have a statistically high co-existence in this population. On the other hand, the Proportional Similarity metric is used for comparing SNP to SNP based on their phenotypic effects. In this case, each vector represents an SNP, with elements being 0 or 1 for associated phenotypes.

%\begin{figure}[h]
%    \centering
%    \includegraphics[width=0.95\linewidth]{fig/comet_image.png}
%    \caption{Similarity and correlation methods in CoMet}
%    \label{fig:comet}
%\end{figure}

%Figure \ref{fig:comet} shows the numerical operations of the two methods.
CoMet has implemented the PS method in both double and single precision using semi-ring GEMM operations, replacing the multiplication in GEMM with \texttt{fminf} operations. Single precision PS was used when the aggregation values were bounded, achieving about a 3.2x higher comparison throughput over double precision.
(The maximum possible speedup is $2\times 5/2 = 5$, where the first factor is of peak performance FP32 over FP64, second factor is the number of instruction involved in protein pair comparison. Because of the modified GEMM and \texttt{fminf} operations used, double precision comparison requires total of 5 instruction per pair, while single precision requires only 2.)

For CCC, CoMet has two different implementations that utilize the fact that the elements in the vectors are 2-bit values. Adopting dense linear algebra ZGEMM operations, it easily separates the 2 bits as separate computations. CCC/bitwise packs 64 vector entries into a double-precision ZGEMM vector but only achieves a 3.6x speedup due to extra bit operations overhead. CCC/tensor\_core, on the other hand, uses two FP16 vectors to hold the two separate bits and performs GEMMex for computation, resulting in a 3x speedup over CCC/bitwise.

The CoMet paper does not quantify the degradation in solution quality from using lower precision for PF and CCC methods, though the result is likely sufficient for downstream tasks. CoMet models all-to-all vector similarities using semi-ring GEMM-like operations, which fall under a FLOP-bounded category that is uncommon for scientific use cases.

CoMet achieved over 6.71 exaflops of FP16/FP32 mixed precision on Frontier \cite{budiardja_caar_2023}.

Ltaief et al. recently carried out GWAS at large scale using mixed precision kernel ridge regression \cite{ltaief_mxp_krr_2024}.
This is one of the few works we reviewed that utilize an FP8 format in the precision mix to deliver performance improvements while controlling accuracy.
The authors use kernel ridge regression to compute associations between a population of patients, Single Nucleotide Polymorphisms (SNPs), some environmental factors and phenotypes such as disease incidence.
This entails
\begin{enumerate}
    \item calculating all-to-all distances and building the kernel matrix $\bld{K}$ from the training genotype matrix $\bld{G}$,
    \item factorizing the regularized kernel matrix $\tilde{\bld{K}}$
    \item solving a linear system from the factorized regularized kernel matrix. and
    \item make predictions on a different population of patients.
\end{enumerate}
Step 1 involves an innovative algorithm to compute pairwise distances between SNPs in the training genotype matrix using matrix-matrix multiplication (GEMM) on tensor cores.
Since the training genotype matrix contains small integer values for SNPs as well as some floating point values, step 1 is carried using mixed INT8 and FP32 GEMMs, selected based on whether the tile consists of integer data or floating-point data. The integer GEMM takes INT8 inputs but accumulates and gives the output in INT32. All these values go into a Gaussian kernel function to generate a symmetric FP32 matrix $\bld{K}$.
Step 2 is carried out by a tile-based adaptive precision Cholesky factorization algorithm using FP32, FP16 and FP8 precision calculations. The precision of each tile is chosen based on its Frobenius norm in relation to that of the whole matrix \cite[sec.~14.3]{higham_mary_2022}.
For the most part, they show that diagonal blocks require FP32 while off-diagonal blocks can be stored in FP8 for the specific datasets considered.
The task runtime system PaRSEC was augmented with precision conversion between tasks of different precision on a block of data.
Step 3 consists of forward- ad back-substitutions (TRSMs) in FP32.

They use their algorithm on a real dataset with 305,880 patients and 43,333 SNPs. In addition, they also use synthetic genetic datasets up to sizes of 13 million patients and 5 million SNPs.
On different supercomputers using NVIDIA A100, GH200 and AMD MI250x GPUs, they show speedups ranging from 3.2$\times$ to 4.8$\times$ faster performance compare to reference FP64 or FP32 computations.
Because of the much higher compute throughput paired with only slightly higher communication performance for low precision on modern hardware, they observe that their mixed precision implementation shows poorer strong scalability than FP64-only computations.
Despite this, at current supercomputers scales, the mixed precision implementation shows the above-stated speedups at large scales on the respective machines.

\subsection{Mixed-precision in AI methods}
\label{sec:aiml}

While not related to a specific application area, we briefly discuss the use of mixed precision in artificial intelligence (AI) applications. Low precision formats (lower than FP32) have been designed with AI training and inference in mind, and scientific applications are increasingly looking to use machine learning (ML) and AI methods along with first-principles calculations.

% Mixed-precision or low-precision AI surrogate models.
Machine-learned surrogate models can augment traditional numerical simulations by replacing some of the steps in a large calculation with machine-learned model. For example, there has been a considerable amount of work in development of machine-learned turbulence models for fluid dynamics \cite{bhushan2023assessment,meena2024machine}, radiation models for climate simulations \cite{pal2019using}, and potential energies for molecular dynamics \cite{nyshadham2019machine}.

AI has also created a new class of applications in data generation (generative AI) and interpolation of large, heterogeneous datasets (data integration).
Data generation can be used to produce candidate solutions to inverse problems.\cite{}  Both are useful for exploring solution space when large datasets are available.

Due to the strong support of deep learning framework for supporting reduced precision calculations, deploying machine-learned surrogates at reduced precision is a rather trivial process which can yield significant gains in performance on NVIDIA's tensor cores and AMD's matrix cores. GPUs such as NVIDIA A100 tensor cores support precision down to 8-bit. 

Deploying machine-learned surrogates at reduced precision generally requires optimizing the neural network for deployment, which generally involves freezing the weights, fusing layers, quantizing the weights, biases, and activations of the model to lower precision, and calibrating the weights (required for INT8 or lower) \cite{brewer2021streaming}. 
While vendors such as NVIDIA boast of 36x improvements in inference performance at reduced precision using their frameworks, deployments in practice of machine-learned surrogates have generally been performed down to 16-bit performance, and have generally have exhibited 2-4x performance speedups \cite{brewer2021streaming} when trained weights are quantized to lower precision off-line.

Studies have been performed to investigate accuracy loss, which generally depend on the type of quantization (e.g., uniform vs. non-uniform), and the performance gains.
Tu et al. \cite{tu_mxp_fno_2024} investigate the use of mixed-precision Fourier neural operator (FNO) blocks for training machine learning models of CFD problems. As opposed to using Pytorch's Automatic Mixed Precision (AMP), this needed some manual implementation to realize the tensor contractions in half precision.
Further, they found that na\"ive application of mixed-precision FNO blocks results in numerical instability due to the reduced dynamic range of the FP16 format. While existing techniques such as loss scaling, gradient clipping etc. did not fix the issue, using \texttt{tanh} activation before each FNO block (`preactivation') restored stability.
The other parts of the model not in the FNO blocks, such as feedforward units, are left to Pytorch AMP.
Over a few different NVIDIA GPUs, they achieve between 1.23$\times$ and $1.58\times$ training throughput.
Using the FP16 FNO block throughout the training results in some loss of accuracy: eg. on a 1024x1024 grid, the error $L^2$ norm with full precision is 0.00213, while that with mixed-precision approach is 0.0026.
The authors additionally propose a precision schedule, where the first 25\% of the training iterations use mixed-precision FNO, the next 50\% use only AMP, and the final 25\% uses full precision.
This technique yields even more accurate training than the full precision baseline, with an error of 0.00170. However, we believe this would reduce the speedup attained.

Wang et al. \cite{wang2024orbit} train a foundation model for earth system predictability using the BF16 format. One of the challenges of training at such precision is that sometimes the gradients cannot be represented within the range that the precision supports. To mitigate such issues, they use PyTorch dynamic gradient scaling to automatically detect and scale the gradients to be within range. 

Sze et al. \cite{sze2017efficient} provide a comprehensive survey of the various techniques used in the efficient processing of deep neural networks, such as network pruning, weight sharing, etc. One of the areas they cover is the impact of reduced precision on energy efficiency. For example, reducing precision from FP32 to FP8 arithmetic can significantly lower energy use in multiply and accumulate (MAC) operations, with energy reductions of 18.5X and 30X respectively.
While these are theoretical per-instruction estimates, researchers have measured real-world energy usage and found significant savings from quantization (use of lower precision) of weights and activations.
Chakravarty \cite{speech_ml_enegy_2024} found, for the Transformer model WhisperX for speech recognition, that the use of FP16 instead of FP32 nearly halves the energy consumption while preserving accuracy.
For Transformer-based time-series models, Kermani et al. \cite{transformer_energy_efficiency_2025} have eperimented with static quantization to INT8 post-training. They find a 29\% reduction in energy consumption while maintaining nearly the same classification performance as FP32.
We should note that this accuracy is about 60\%. 
% What's the performance benefit to the overall application?
% -- Let's echo this question to the reader during the Conclusion section

\subsection{Observations from the application areas}
\label{sec:appl_conclusions}

We now make some observations which are common across several of the reviewed application areas.

\begin{enumerate}
    \item Mixed-precision algorithms may need to be tailored to the specific physical modeling application and overall numerical method within which the mixed precision solver works.
    \item The use of lower precision at inner levels of numerical solvers, such as the preconditioner in Krylov solvers \cite{appl_wc_ackmann} or the linear iterations in a nonlinear solver \cite{walden_unstructured_mixed_2019}, is a recurring theme.
    \item Some projects can use lower precision in only some regions of the spatial domain \cite{appl_wc_ackmann} because using low precision everywhere results in unacceptable errors. Similarly, some science areas use other domain-specific parameters to determine where to use low precision \cite{titov_quantum_2013} and determine subspaces in which the solution contribution tends to zero as the iterations progress \cite{das_dftfe_2019}.
    \item There are often input uncertainties to a model, such as uncertainties in initial conditions or inaccuracy in domain-specific, approximate sub-models.
    In many cases \cite{saffin_reduced-precision_2020, appl_wc_ackmann}, such models can accept additional error from using mixed precision and still provide solutions with uncertainty comparable to the input uncertainty.
    This is also applicable when it is statistics of the solution which are physically relevant rather than individial instances of the solution \cite{karp_mxp_scale_resolving_2025}.
    \item Additive transformations of variables, such as conversion of temperature from K to $^\circ C$ or expressing a variable as a correction with respect to some reference values \cite{saffin_reduced-precision_2020}, can help move the range of realized values to a region where the relative error is the lowest.
    \item Significand bits (\emph{precision}) are important in scientific applications, not just exponent bits (range) \cite{saffin_reduced-precision_2020}.
\end{enumerate}

We have not commented on the energy efficiency implications of used mixed- and reduced-precision techniques in computational science applications, but it must be noted that reduced precision implementations can enable significant improvements in energy efficiency.
For example, Sakamoto et al. \cite{sakamoto2020effectiveness} studied the effect of low-precision computing on energy efficiency for two target applications: (1) solving Poisson's equations using the preconditioned conjugate gradient method with incomplete Cholesky factorization (ICCG), and (2) earthquake ground vibration simulation. They were able to achieve 34\% and 38\% reduction in energy consumption in each application respectively by going from double precision to single precision. They mention that using mixed precision offered the best balance of both reducing energy consumption and also maintaining sufficient numerical accuracy. 

% \begin{table}
%     \centering
%     \begin{tabular}{|c|c|c|}
%     \hline
%     \textbf{Appl. domain} & \textbf{Limiting resource} & \textbf{Speedup} \\
%     \hline
%     Aerodynamics         & Mem. BW        & 1.44* \\
%     Other CFD            & Mem. BW        & 1.6*  \\
%     Weather and climate  & Mem. BW*       & 1.4 \\
%     Sequence similarity  & Compute        & 3.6 \\
%     \hline
%     \end{tabular}
%     \caption{Representative speedups obtained from mixed-precision methods in different application areas. Asterisks represent the use of some reasonable assumptions by the current authors based on the literature.}
%     \label{tab:appl_speedups}
% \end{table}

% Please add the following required packages to your document preamble:
% \usepackage{booktabs}
\begin{table}[h]
\centering
\begin{tabular}{@{}lll@{}}
\toprule
\textbf{Application Domain} & \textbf{Resource Bottleneck} & \textbf{Speedup} \\ \midrule
Aerodynamics                & Mem. BW                      & 1.44*            \\ \midrule
Other CFD                   & Mem. BW                      & 1.60*             \\ \midrule
Weather \& Climate          & Mem. BW*                    & 1.40              \\ \midrule
Quantum chemistry           & Mixed                        & 1.9*            \\ \midrule
Computational genomics      & Compute                      & 4.80             \\ \midrule
%AI Model Training      & Compute                           & 1.25-1.58        \\ \midrule
%AI Inference      & Compute                                & 2-4              \\ \bottomrule
\end{tabular}
\caption{Representative speedups obtained from mixed-precision methods in different application areas. Asterisks represent the use of some reasonable assumptions by the current authors based on the literature.}
\label{tab:my-table}
\end{table}

\section{State of the art in mixed-precision numerical algorithms for science}

We now turn our attention to the fact that mixed-precision algorithms have been formulated for some computational \emph{motifs} that recur in computational science.
Notably, the US Department of Energy has made some investments in mixed-precision numerical methods as part of the Exascale Computing Project (ECP) via the xSDK mixed-precision effort.
The main participants of the effort have documented \cite{ecp_mxp_advances_2021} progress in dense linear solvers, sparse direct solvers, sparse iterative solvers, sparse approximate inverse preconditioning, multigrid methods and discrete Fourier transforms.
For the broad field of numerical linear algebra, Abdelfattah et al. \cite{abdelfattah_survey_2021} and Higham and Mary \cite{higham_mary_2022} provide a greater survey of mathematical results related to mixed-precision algorithms.

%Here, we focus on the sparse computations and Fourier transforms as these are relevant for science and engineering simulations.
Dense matrix factorization is one of the oldest motifs in computational science. The primary benchmark for measuring supercomputer performance, High Performance Linpack \cite{netlib_hpl}, measures the time taken to complete a large distributed dense matrix factorization into triangular matrices.
This operation, known as $LU$ factorization, is one of the primary problems in applied mathematics and numerical methods.
Certain problems in quantum chemistry make heavy use of dense matrix operations, as described in section \ref{area:quantum}.
However, dense factorization of an $n\times n$ matrix requires $\mathcal{O}(n^3)$ arithmetic operations.

In many scientific fields, sparse matrices are used. These include computational fluid dynamics, fusion plasma simulations, weather simulations, electromagnetics, macro-scale and multi-scale material science (phase field modeling, plasticity, crack propagation etc.) and many others.
Factorization of square sparse matrices into triangular factors (sparse $LU$ factorization) using Gaussian elimination is a robust, cost-efficient and memory-efficient method to solve ill-conditioned large sparse linear systems arising, for example, in power grid simulations \cite{swir_power_opt_linear_solvers}.
These are called sparse direct solvers.
On very regular grid topologies with good ordering of the unknowns, sparse direct methods can solve the problem in $\mathcal{O}(n^{3/2})$ arithmetic operations, though in practice the scaling is somewhat worse \cite{george_ND_1973,george_1WD_1980}.

For problems which are not too ill-conditioned, this scaling can be improved upon by iterative methods for sparse linear systems.
Iterative methods compute an approximate solution to the linear system by starting with an initial guess and successively improving it based on the principles of projection (into a low-dimensional subspace) and \emph{preconditioning}.
Iterative solvers incorporate inner operations called preconditioners which attempt to modify the problem such that the solution remains the same, but it gets easier to solve. Note that the preconditioner may itself be another iterative solver.
With a carefully-constructed preconditioner well-suited to the scientific domain, parallel iterative solvers can even achieve the holy grail of $\mathcal{O}(n \log n)$ time to solution, thus being very scalable.
This is enabled, in particular, by multigrid methods \cite{brandt_multigrid_1977}. These are special preconditioners that construct a hierarchy of smaller (coarser) problems that compensate for the inaccuracies of traditional iterations on the original large (fine) problem.
In addition to multigrid methods, there are a variety of simpler preconditioners that are favoured in different classes of simulations.

Apart from multigrid, another computational motif that provides for very scalable solutions to certain kinds of problems is the Fast Fourier Transform (FFT). This is a class of scalable $\mathcal{O}(n \log n)$ algorithms to compute the discrete Fourier transform of a function discretized at $n$ points in space or time.

The motifs described above are \emph{linear} operations (or approximations of them) in a mathematical sense. Frequently, they arise in the context of computing solutions to nonlinear problems, including nonlinear differential equations.
A common computational motif in this regard is Newton's method to solve (find a zero of) a system  of nonlinear equations.
For robustly computing solutions to nonlinear problems, it is necessary to study mixed-precision variants of motifs like Newton's method.
Finally, we also consider eigenvalue problems and ordinary differential equations (ODEs) which are important in a variety of scientific fields, including quantum chemistry, CFD etc.

In this section, we review the existing state of the art in mixed-precision algorithms for these motifs.
We discuss the attained speedups and any accuracy degradation.
\begin{greenbox}
    A key idea in mixed precision algorithms is that often, when a calculation deals with values with widely varying magnitudes, values with small magnitudes can be effectively represented in low precision \cite[sec.~14]{higham_mary_2022}.
\end{greenbox}

Iterative refinement (IR) \cite{moler_ir_1967} is a starting point for many of the techniques to calculate sufficiently accurate solutions while using lower precision.
For solving a linear system of equations, the algorithm is shown in listing \ref{alg:base_ir}.
In the case of dense linear systems, $\bld{G}$ is typically the application of a pre-computed LU factorization in low precision.
\begin{algorithm}
    \begin{algorithmic}
        \Require Initial guess $\bld{x}_0$ in working precision $u$.
        \While {$i < N$}
            \State Compute residual $\bld{r} \gets \bld{b} - \bld{A}\bld{x}_i$ in (high) precision $u_r$.
            \State Apply procedure $\Delta\bld{x} \gets \bld{G}(\bld{r})$ using one or more lower precision formats.
            \State $\bld{x}_{i+1} \gets \bld{x}_{i} + \Delta\bld{x}_i$ in working precision $u$.
        \EndWhile
    \end{algorithmic}
    \caption{General iterative refinement for linear system $\bld{Ax}=\bld{b}$}
    \label{alg:base_ir}
\end{algorithm}

\subsection{Matrix multiplication}

One can approximate the fundamental operation of matrix multiplication using lower-precision compute units and formats with a known and controlled error.
Typically, the specific operation targeted is dense matrix-matrix multiplication due to the hardware units of interest being tensor cores.
These can be incorporated into BLAS libraries and be used within solvers to transparently gain the advantages of the lower precision compute units.
This type of approach is also referred to as `emulation', though they do not literally attempt to emulate IEEE FP64 arithmetic.
A common class of techniques that has recently been proposed is that of splitting schemes, where high-precision matrices are split into a sum of several lower-precision components. Notable approaches include multiword arithmetic and the Ozaki split schemes (discussed in subsections below).
As GPU architectures evolve, driven by AI and machine learning demands, these splitting techniques continue to advance, aiming for more efficient and accurate approximations of high-precision computations.
Independent of `emulation' type methods, at least one method has also been proposed for sparse matrix vectors products.
These are fundamentally based on the idea that numbers with relatively small magnitude can be represented in narrower formats without catastrophic rounding errors.
We briefly survey these methods below.

%So far, we have described numerical methods that use lower precision in sophisticated ways to solve linear and nonlinear systems of equations.

\subsubsection{Multiword arithmetic}
\label{sec:multiword}
Higham and Mary \cite[section~13]{higham_mary_2022} give a description of multiword arithmetic, and Fasi et al. \cite{fasi_multiword_2023} developed a multiword based mixed-precision GEMM that utilizes GPU tensor cores. 
The idea behind multiword GEMM is to split the matrix based on the remainder with respect to a lower precision representation.
\begin{greenbox}
Given two matrices $A \in  \mathbb{R}^{m\times n}$ and $B \in  \mathbb{R}^{n\times q}$, $C=AB$ is estimated by
\begin{equation*}
C' = A_1B_1 + A_1B_2+ A_2B_1,
\end{equation*}
where $A_1$, $B_1$ is the lower precision representation of $A$, $B$, and $A_2$, $B_2$ is the difference between two precision (eg. $A_2=A-A_1$).

This example constitutes 2-word arithmetic. Generalizing to $p$-word arithmetic, the matrices can be split into $p$ approximate matrices each, increasing the precision of the result at the cost of increased low-precision matrix multiplications.
\end{greenbox}

Fasi et al. reported \cite{fasi_multiword_2023} a speed up of 2.2x and 7.3x over the SGEMM (FP32) on V100 and A100, by using FP16 representation of $A_1$ and $B_1$. The theoretical error bound was also analyzed: in theory, multiword arithmetic has the same accuracy as high-precision arithmetic. However, when na\"ively applied on GPUs, the magnitude of error is significantly larger than the higher precision arithmetic due to the rounding mechanism of the GPU.
FABsum was proposed in \cite{fasi_multiword_2023} to eliminate this rounding issue by changing the inter-block accumulation to higher precision.

We point out that multiword arithmetic can be used not only for dense GEMM, but also in other motifs and applications such as Fast Fourier Transform (section \ref{sec:fft} and sparse matrix operations in weather simulations (section \ref{sec:appl:weather}.

In 2022, a scheme based on multiword arithmetic was applied by Ma et al. \cite{ma_emulation_2022} to accelerate dense matrix-matrix multiplication as well as convolution.
This work favors approximating FP32 operations at the scalar level by defining new datatypes made up of FP16, TF32, INT16 (fixed point) or BF16 numbers.
It also introduces a general framework to select the precision based on error.
An example is their FP32-F type, which consists of two FP16 values. The FP32 multiplication is then emulated by three FP16 multiplications and one addition. The overall performance of FP32-F GEMM was 3.12x, 1.81x, and 1.49x faster than \texttt{cublasSGEMM} on NVIDIA A100, V100, and T4, respectively.
Additionally, the paper shows that the maximum relative error for specific use cases falls within a magnitude of SGEMM.

\subsubsection{Ozaki Splitting}
\label{sec:ozaki_1}
The Ozaki split scheme was first introduced in 2011 \cite{ozaki2012error}. The split scheme's original purpose was to support higher precision computing by splitting the floating-point number into several floating-point numbers to increase the size of significant bits.
This is often used for FP32 to emulate the accuracy of FP64 on consumer-grade hardware that does not have physical FP64 compute units.

In 2020, Mukunoki \cite{mukunoki2020dgemm} adopted the Ozaki split scheme and implemented DGEMM (double precision matrix-matrix multiplication $C = A\times B$) on tensor cores.
\begin{greenbox}
\begin{enumerate}
\item The input matrices $A$ and $B$ are split into two sets of multiple FP16 matrices $S_A$ and $S_B$.
\item FP16 GEMMs are carried out for those split matrices \[ S_{Ci}=\sum_jS_{Ai}S_{Bj}. \]
\item the final output is calculated by aggregating the multiple FP16 output matrices, $C=\sum_iS_{Ci}$.
\end{enumerate}
The multiplication carried out on these FP16 matrices was placed on the tensor core, and the number of splits was determined by the value range in the input matrices and the precision required.
\end{greenbox}

The results in this paper show that the accuracy of this emulation is actually higher than \texttt{cublasDGEMM}, and the performance is higher than \texttt{cublasDGEMM} on accelerators (Titan RTX) that do not have physical double precision computing units.
However, when compared to accelerators (Tesla V100) that have physical double precision computing units, this method is about 6.7x slower than \texttt{cublasDGEMM}.
Furthermore, this method requires additional memory to store the split matrices.
For example, for square matrices with sizes between 2048 and 4096 and a particular range of values, computing GEMM with accuracy equivalent to FP64 requires 10 splits stored in FP16.
In this case, this technique would require $10/4\times$ or 2.5$\times$ the memory that a regular FP64 GEMM would need.
While this factor can be brought down, to an extent, by blocking along the outer GEMM dimensions, it is a good ballpark estimate.

%Similar efforts have been made, with similar results,
FP64 matrix multiplication has also been demonstrated using INT8 split matrices on integer tensor cores \cite{ootomo2024dgemm}.
The intuition behind emulating in integer tensor cores is that they have more arithmetic operations per second than FP16 tensor cores while still able to maintain the same accuracy.
This method has achieved higher throughput than direct FP64 GEMMs on consumer-oriented GPUs.
However, it is a few times slower than direct DGEMM calls on NVIDIA A100, which contains physical double precision tensor cores.

In Ozaki splitting, the number of required splits, and thus the number of lower precision GEMMs, depends on the input matrix's range of values.
The wider the range, the more the number of splits required to emulate GEMM in FP64 precision. 
In their experiments \cite{ootomo2024dgemm}, the input matrix is randomly generated and their magnitudes are distributed based on a parameter $\phi$.

\begin{greenbox}
Input matrices are generated using the distribution
\begin{equation}
\mathcal{U}(-0.5,0.5)\times e^{\phi \times \mathcal{N}(0,1)},
\label{eq:ozdist}
\end{equation}
where $\mathcal{U}$ and $\mathcal{N}$ refer to uniform and normal distributions respectively.
For example, a matrix with $\phi=1$ has a value range of $1.4\times 10^{-9}$ to $1.6\times 10^{2}$.
\end{greenbox}

Preliminary work at NVIDIA has leveraged this class of techniques to achieve remarkable gains in energy efficiency on their latest Blackwell hardware [private communication].
These include gains in solving dense linear systems in the High-performance Linpack (HPL) benchmark, where a 1.8$\times$ advantage was achieved in compute throughput per watt of power. They also claimed that this efficiency gain carries over to important scientific applications that depend on dense GEMMs. 
%\begin{figure}
%    \centering
%    \begin{subfigure}{0.44\linewidth}
%        \includegraphics[width=0.99\linewidth]{fig/nvidia-emulated-dgemm-hpl-cropped.png}
%        \caption{DGEMM and HPL}
%        \label{fig:gb200-dgemm-hpl}
%    \end{subfigure}
%    \begin{subfigure}{0.55\linewidth}
%        \includegraphics[width=0.99\linewidth]{fig/nvidia-emulated-apps.png}
%        \caption{Scientific applications}
%        \label{fig:gb200-apps}
%    \end{subfigure}
%    \caption{Energy efficiency advantages from an Ozaki-like approach on recent NVIDIA GPUs. Preliminary results, courtesy NVIDIA.}
%\end{figure}
%For these scientific applications from computational chemistry and material science, performance per watt improvements ranging between 4$\times$ to 34$\times$ were obtained; see Figure \ref{fig:gb200-apps}.

\paragraph{Is the Ozaki scheme currently relevant for science?}

Our review above indicates that current data center accelerators, which include dedicated physical FP64 computing units, would not see the benefit of this type of approximation commonly referred to as `emulation'.

In order to see performance improvement over FP64 GEMM using emulation on low-precision tensor cores, they need to have roughly $100\times$ peak FLOPS over the double precision units. We estimate this approximate threshold using results from Ootomo et al.\cite{ootomo2024dgemm}.
Their $\phi=0.5$ matrix would require 8 splits and the $\phi=4$ matrix would require 11 splits, which is equivalent to 36 and 66 INT8 GEMM operations respectively. Therefore, in theory, assuming the GEMM efficiency remains the same across levels of precision and there is no split overhead, the threshold is at $\frac{INT8 OPS}{FP64 FLOPS} >$ 36 to 66. 

In practice, the GEMM efficiency and the split overhead must be considered. In the paper \cite{ootomo2024dgemm}, the authors have observed that the split overhead (including accumulation) is around 20\% for a square matrix with size up to 16K. %This ratio will decrease as the matrix size becomes larger.
Furthermore, the paper reports that the DGEMM on NVDIA A100 using a tensor core achieves 17.5 TFLOPS (90\% hardware peak) and the INT8-GEMM only achieves 375 TOPS (60\% hardware peak).
Putting these two additional factors together, our estimate is that the hardware's low-precision units must have approximately $81\times$ to $118\times$ maximum FLOPS over the double precision units depending on $\phi$, with current software on the NVIDIA A100.
With more recent optimizations, the authors believe that a minimum of 40 to 60× performance over FP64 units is required to achieve speedup [private correspondence].

We see a positive impact of emulation in the future on two fronts.
First, upcoming AI-focused hardware is beginning to have $\frac{TC_{low} OPS}{FP64 FLOPS} > 100$, or in the future they may not even have physical double precision units.
Such hardware could potentially rely on these emulation methods for double precision calculation.
For example, Dongarra et al. \cite{dongarra_hardware_trends_2024} claim that using the latest version of Ozaki splitting \cite{uchino_ozaki_2024}, they achieve 2$\times$ speedup over FP64 using 7 splits. We point out here that while 7 splits is enough for the HPL benchmark, it will not achieve double precision accuracy in all cases. Regardless of this caveat, `emulation' is promising on the latest NVIDIA hardware.
Second, a science domain that has a matrix with a narrower value range could potentially benefit from this emulation, as such a matrix would require fewer splits and smaller ratios. For example, the HPL benchmark appropriately passes its residual check with only 7 splits, which is equivalent to only 28 INT8-GEMMs.

% In order to see performance improvement of FP64 GEMM using emulation on low precision tensor cores, they need to have roughly 100x peak FLOPs over the double precision units. The estimate of this turning point is calculated based on
% the observed slowdown of emulation $\times$ current $\frac{TC_{low} FLOPS}{FP64 FLOPS}$.
% Since the number of splits required depends on the precision needed, achieving FP64-equivalent accuracy may necessitate more than 8 splits on FP16 and 10 splits on INT8, which in turn demands a performance of at least 100× over FP64 units. We interviewed the authors, and with more recent optimizations, they believe that a minimum of 40 to 60× performance over FP64 units is required to achieve usefulness.
% However, new emerging AI-focused hardware may have $\frac{TC_{low} FLOPS}{FP64 FLOPS} > 100$, or may not even have physical double precision units.  Such hardware could potentially rely on these emulation methods for double precision calculation.

% Hao
To further enhance the real-world applicability of the Ozaki scheme in scientific computing, Abdelfattah et al.~\cite{abdel_ozaki_analysis_2025} conducted a comprehensive error analysis of basic GEMM and LU factorization across several types of matrices. They introduced a method for determining the number of required splits based on the dynamic range of the input matrix. This approach relies on analyzing the bit-level representation of matrix entries.

\begin{greenbox}

\begin{enumerate}
    \item Compute the \textbf{bit spread} $\delta_{ij}$ by identifying the positions of the most significant and least significant bits in the mantissa of each element $a_{ij}$ of the matrix $\bld{A}$:
    \[
    \delta_{ij} = \mathtt{msb\_idx}(a_{ij}) - \mathtt{lsb\_idx}(a_{ij})
    \]
    
    \item Extract the \textbf{exponent} $e_{ij}$ from the binary representation of each value:
    \[
    e_{ij} = (a_{ij} \gg 52) \,\&\, \mathtt{0x7FF}
    \]
    
    \item Estimate the required number of \textbf{splits} $s$ as:
    \[
    s = \frac{ \max_{i,j} \delta_{ij} + \left( \max_{i,j} e_{ij} - \min_{i,j} e_{ij} \right) }{7}
    \]
\end{enumerate}
\end{greenbox}

This formulation enables the estimation of the minimum number of components required to emulate high-precision arithmetic using low-precision hardware.

% Rogers: I added \begin\end greenbox tags and moved the subsection header below into a box to see how it looks.
% This was suggested as a way to focus the paper on mixed precision use patterns - by putting instances of applying the pattern into call-out boxes.
%\subsection{Ozaki scheme II}

%\label{sec:ozaki_ii}
\paragraph{\textbf{Ozaki scheme II}}

Recently, Ozaki et al. \cite{ozaki_ii_2025} proposed an alternative emulation scheme based on the Chinese Remainder Theorem with greatly improved performance and memory footprint.

\begin{greenbox}
In the Ozaki II technique, pairwise coprime integers $m_i$, $i=1,...,s$ are chosen in order to express an INT8 number $a$ uniquely as a vector of $s$ numbers $a_i$:
\begin{equation}
    a \equiv a_i \mod m_i, \quad i = 1,...,s.
\end{equation}
According to the Chinese Remainder Theorem, if the product $M:=\Pi_{i=1}^s m_i$ is large enough, the original number $a$ can then be reconstructed from the $a_i$ using
\begin{equation}
    a \equiv \sum_{i=1}^{s} a_i \frac{M}{m_i} y_i \mod M
\end{equation}
where $y_i$ is defined by $M_iy_i \equiv 1 \mod M$.
This method is applied to matrix multiplication of integer matrices $\bld{A}' \in \mathbb{Z}^{p\times q}$ and $\bld{B}' \in \mathbb{Z}^{q\times r}$. If
\begin{equation}
    \bld{C}_i \equiv \bld{A}'\bld{B}' \mod m_i
\end{equation}
where the equivalence holds elementwise, then the product $\bld{C}$ is given by
\begin{equation}
\bld{C} \equiv \bld{A}'\bld{B}' \mod M = \sum_{i=1}^s \bld{C}_i \frac{M}{m_i}y_i \mod M.
\end{equation}

Double-precision matrices are quantized to integer matrices using a method similar to that used by Ootomo et al. \cite{ootomo2024dgemm}, though the scaling factors now depend on the product $M$ of the pairwise coprime moduli.
\end{greenbox}

Note that as opposed to the Ozaki scheme I of the previous section, this method only needs $s$ INT8 GEMMs for $s$ moduli (see box). This becomes $s+1$ INT8 GEMMs in practice, including one to estimate the largest entry in the product.

Again, drawing real matrix entries from the distribution \eqref{eq:ozdist}, Ozaki et al. \cite{ozaki_ii_2025} show that 14-18 moduli are sufficient for attaining double-equivalent accuracy for $\phi \in \{0.5,1,2,3,4\}$.
Recall that the Ozaki scheme I requires about 8-11 splits for double-equivalent accuracy. In the favorable case of 8 splits, this requires $s(s+1)/2$ INT8 GEMMs, which are 36 to 66.
Meanwhile, Ozaki scheme II requires only 15-19 INT8 GEMMs, significantly reducing the runtime.
Furthermore, in theory, the new scheme can be optimized so that there is a fixed constant smaller memory overhead independent of $s$, as opposed to the original scheme.

With the Ozaki scheme II (see box), Ozaki et al. \cite{ozaki_ii_2025} achieve emulated performance even faster than native DGEMM performance on the NVIDIA GH200 chip. For 16,384 size square matrices, they attain 71.1 TF using 14 moduli, while native DGEMM using cuBLAS gives 60.9 TF.
This is a ground-breaking result that, if reproduced across different real-world matrices, may revolutionize GPU architecture.

\subsubsection{Sparse matrix vector product}

Graillat et al. \cite{graillat_adaptive_spmv_2024} have proposed an adaptive precision sparse matrix vector product algorithm, along with error analysis.
\begin{greenbox}
Using $q$ precision formats with unit roundoffs $u_1 < u_2 < ... < u_q$, they propose partitioning each row of a sparse matrix $\bld{A}$ into `buckets' $B_{ik}$ containing column indices, $k \in \{1, ..., q\}$.

The $k$th partial inner product $\hat{y}_i^{(k)} := \sum_{j \in B_{ik}} a_{ij}x_j$ is computed in precision $u_k$.
These buckets can perhaps be compared to slices in the Ozaki scheme.

Grailat et al. show that the norm-wise backward error (as opposed to the component-wise backward error) depends on the ratio
\begin{equation}
    \frac{\sum_{j\in B_{ik}}|a_{ij}|}{\lVert \bld{A} \rVert}.
\end{equation}
In this framework, setting $u_q = 1$ also gives a robust dropping strategy. In the norm-wise accuracy approach, if $\epsilon$ is the desired norm-wise relative error, then entries smaller than $\frac{\epsilon\lVert A\rVert}{u_q}$ may be dropped. They demonstrate experimentally that this dropping is frequently a major contributor to the speedups obtained from the method.
\end{greenbox}

They test their algorithm over a range of 32 sparse matrices from the Suitesparse collection as well as other matrices from numerical simulations.
They run these tests on an Intel Sylake CPU using an OpenMP implementation which is not optimized. However, claimed benefits should extend to GPU-based system as they primarily come from reducing the memory traffic on a memory bandwidth limited kernel.
Using two precision formats, FP64 and FP32, while guaranteeing a norm-wise accuracy of $2^{-53}$, they show performance impact ranging from a slowdown of about 0.91$\times$ to speedup of 4$\times$.
The method can also reduce memory footprint by up to $6\times$ while achieving $2^{-53}$ accuracy in the norm-wise sense.
Both the speedup and memory benefits increase if we can decrease the accuracy requirement. For example, if we increase the normwise tolerance to $2^{-37}$, the speedup is as much as $7\times$ and the memory footprint drops by as much as $10\times$.

Further, they also tested their adaptive SpMV algorithm within iterative linear solvers such as GMRES and BiCGStab. Using GMRES-IR (40) with a simple Jacobi preconditioner, they observe speedups between 1.1$\times$ and $6\times$ depending on the matrix properties.
The upper end of the speedups results primarily from the fact that sometimes, the adaptive precision representation with dropping becomes much better conditioned and leads to convergence in much fewer iterations.

\subsection{Dense factorizations}
\label{sec:dense_factorization}

Mixed precision iterative refinement (IR) has been suggested for dense matrix factorization and solution for many years.
Haider et al. \cite{magma_mxp_2018} gives a detailed study of its performance on different kinds of matrices, number formats and IR variants. They carry out LU factorization with partial pivoting using low precision and attempt to recover FP64 accuracy by iterative refinement.
They find that IR with FP16 factorization precision using tensor cores is generally a good option for matrices with positive eigenvalues bounded away from zero.
When the matrix has mixed eigenvalues, especially when the singular values are not clustered, IR generally requires more iterations and may even stall.
There are some cases where GMRES-IR succeeds but LU-IR is very slow or diverges.
On a Tesla V100 GPU, they achieve a 4$\times$ speedup over double-precision LU factorization for diagonal-dominant matrices. In their worst case scenario for indefinite matrices with about $10^5$ condition number, they see a 3$\times$ speedup.
The tensor-core version of the FP16 algorithm is numerically more suitable compared to not using tensor cores, since it accumulates the matrix sum in FP32 arithmetic.

HPL-MxP is the first benchmark that designed to explore the maximum mixed-precision capability of a system.
The problem statement consist with a single dense linear solve using mixed-precision LU factorization and a double-precision iterative refinement (IR).
The matrix used in this benchmark is strictly diagonal-dominant for removing the partial pivoting requirement and for reducing the number of IR iterations.
Hence, this benchmark is only limited by raw computational power (FLOPs bounded) until the trailing matrix size is too small for overlapping communication and computation. 
The pseudo-code of HPL-MxP is shown in the box below. The precision is indicated in the front of each numerical operation.

\begin{greenbox}
%\begin{algorithm}
    \begin{algorithmic}
        \Require Initial guess $x$ in double precision, block size $n_b$.
        \State $A$ = FP32\_CAST ( \textbf{A} )
        \While {$N_i < N$} \Comment{factorization loop}
            \State (FP32) $A = A$[$N_i$ : , $N_i$ : ] 
            \State (FP32) Triangular factorization $L_{1,1}U_{1,1}$ = $A_{1,1}$ 
            \State (FP32) Broadcast $L_{1,1}$ and $U_{1,1}$ 
            \State (FP16) Tri. solve $L_{2,1} = {A}_{2,1}U_{1,1}^{-1}$, $U_{1,2} = L_{1,1}^{-1}A_{2,1}$  
            \State (FP16) Broadcast $L_{2,1}$ and $U_{1,2}$ 
            \State (FP16) Update $A$ = $A-L_{2,1}$$U_{1,2}$
            \State \indent $N_i += n_b$
        \EndWhile
        \State (FP64) $x$ = Iterative Refinement ($A$, \textbf{A}, $b$).
    \end{algorithmic}
    %\caption{HPL-MxP solving $\textbf{A}x=b$}
    %\label{alg:lu_ir}
%\end{algorithm}
\end{greenbox}

Due to its compute-bound nature, HPL-MxP is expected to speed up the double precision counter part (HPL) by the theoretical factor of mixed-precision hardware capability, i.e., the ratio of speed-up is expected to be close to hardware's ratio of FP16 to FP64 FLOP throughput.
Most of the speed up compared to the HPL benchmark is achieved by reducing the precision of the off-diagonal blocks, both in solving and updating.
The HPL-MxP implementation of Lu et al. \cite{hpl_mxp_2022} achieved a speed-up over HPL with the ratio of 9.50x for Summit and 8.31x for Frontier. The hardware FLOPs ratio for NVIDIA V100 is 16.02x and for AMD MI250x is 7.99x.
The reason the observed speedup is lower than the hardware ratio is due to several reasons.
(1) The theoretic peaks reported by hardware vendors are not the actual achievable peaks of their BLAS libraries.
(2) There are still optimizations that were left out for the HPL-MxP implementation on Summit.
(3) GPU memory capacity limited the problem size especially on Summit.
(4) HPL-MxP still requires some FP32 precision operations during the execution.
These facts limit the achievable speed-up.
Nevertheless, the technology developed can be used to achieve very high performance on scientific applications that require large-scale dense matrix factorization.

In the case of Cholesky factorization for symmetric positive definite matrices, Abdelfattah et al. \cite{abdelfattah_symmetric_fp16_2020} showed that 4.7$\times$ speedup can be observed depending on the matrix eigenvalue distribution.
The notable fact here is that they developed a preprocessing stage that scales the matrix and shifts the diagonal if necessary in order to maintain positive definiteness in low precision.

\subsection{Sparse factorization and linear solvers}

Similar to dense linear solvers, mixed-precision iterative refinement (IR) has been used for sparse direct solvers in SuperLU\_Dist using FP32 factorization and FP64 residual computations and updates.
Unlike for dense solvers, the update step in sparse solvers is not dominated by GEMM (matrix-matrix multiplication). In addition, there are gather and scatter operations owing to the sparse structure of the matrix \cite{ecp_mxp_advances_2021}.

There was some early work on various techniques to accelerate sparse linear solvers by mixed-precision techniques by Buttari et al. \cite{buttari_mxp_sparse_64bit_2008}.

More recently, on some Suitesparse matrices \cite{suitesparse_coll}, the use of tensor core GEMM calls led to less than 5\% speedup in the whole sparse LU solver \cite{ecp_mxp_advances_2021}. However, when comparing FP64 and FP32 version, about 50\% speedup was still observed.
The iterative refinement is usually less than 10\% of the factorization time, and therefore, overall, the mixed-precision IR implementation is faster than using FP64 everywhere.

Amestoy et al. studied mixed-precision iterative refinement with sparse approximate factorization more recently \cite{amestoy_sparse_ir_2023}.
Approximate algorithms they consider for factorization include block low rank methods and static pivoting (instead of dynamic partial pivoting).
In block low rank methods, individual blocks are represented adaptively by a truncated SVD thereby compressing them. Static pivoting refers to replacing the traditional partial pivoting operation, which is latency and communication intensive, with just replacing very small values by a fixed constant dependent on the matrix norm.
The authors combine the use of these approximations with the use of low-precision arithmetic as well as iterative refinement loops around it to achieve the desired accuracy.
Since sparse factorization suffers from fill-in (the $L$ and $U$ factors are denser than the original matrix $A$), the use of low precision to store the factors not only speeds up computations but also reduces memory consumption.
They note that the residual computation $r = b - Ax$ in IR has a very low cost compared to the triangular solves. This means a higher precision can be used for the residual, thus attaining higher overall accuracy, while not increasing the overall execution time significantly.
The authors carry out an error analysis, where the main take-ways are:
\begin{itemize}
\item LU-IR converges only if GMRES-IR converges, and the latter may converge in cases where the former does not.
\item When using approximations that lead to large element growth during factorization, it may be necessary to use high precision for stability, even though the approximation is itself reducing the accuracy to a degree.
\end{itemize}
They carry out experiments with Suitesparse matrices and some other matrices from cardiac electrophysiology, incompressible automobile aerodynamics, noise propagation, structural mechanics of automobiles and turbomachinery, electromagnetics as well as a coupled multiphysics problem.
All computations were done one 1 or 2 nodes of a dual-socket Skylake CPU system.
Hardware and software support for half precision was still an issue on CPUs; they do not investigate its use.
The key experimental findings are:
\begin{itemize}
    \item GMRES-IR typically requires more $LU$ solves than the corresponding LU-IR and, in their case, more memory due to cast copies. ($L$ and $U$ are case to higher than working precision in the case of GMRES-IR.) However, GMRES-IR is typically more robust - it succeeds in some cases where LU-IR fails, as expected from the error analysis.
    \item In those cases where mixed-precision LU-IR fails but GMRES-IR succeeds, the latter usually needs more time than a classical high-precision factorization though it uses less memory.
    \item Utilizing quadruple precision for residual computations but single precision for factorization is typically faster than classical double-precision factorization. Quad-single GMRES-IR is generally reliable while accuracy of quad-single LU-IR is sometimes poor.
    \item In favorable cases, when the factorization time dominates compared to IR time, speedups between 1.33$\times$ and 1.6$\times$ can be observed for FP32 factorization and FP64 residuals.
    \item For other problems, however, the use of mixed-precision IR can have results ranging from a speedup of 1.25$\times$ to a slowdown of 0.57$\times$.
    These problems consist of those which are too small, those that produce little fill-in, and those that are badly conditioned and require too many IR iterations. In all three cases, the cost of all BLAS-2 operations is not small compared to factorization.
    \item Mixed-precision IR always utilizes less memory than FP64 factorization and solve. The gains can vary from 2$\times$ to 1.14$\times$.
    \item Sometimes, it is possible to find solver parameters to trade-off some slowdown in time-to-solution for significant memory savings.
\end{itemize}
The potential downsides of using mixed-precision IR for sparse factorization are as follows.
\begin{itemize}
    \item  The analysis phase of the factorization, where the sparsity pattern of $L$ and $U$ is computed, uses only integer operations and mixed-precision is not applicable to this. This reduces the achievable speedup from using mixed-precision.
    \item With the many different precision formats involved, casting operations may become an overhead in terms of time or memory. Eg., GMRES-IR potentially requires the $L$ and $U$ factors in different precision from the one in which they were computed.
    \item For some problems, integer data and operations can meaningfully reduce speedups obtainable from the use of lower-precision floating point types for numerical data. This includes the sparse storage format such as those needed for compressed sparse row (CSR) storage, manipulating index sets for reordering and partitioning, etc.
\end{itemize}

Zounon et al. \cite{zounon_sparse_direct_mxp_2022} experimentally analyze the performance of different sparse direct solver packages when using single precision factorization with iterative refinement and find that many problems can see a speedup of $1.5\times$ or above, but for other problems this is compromised by either encountering subnormals in FP32 which cause slowdowns, or expensive analysis phases that do not benefit from lower precision.

\subsection{Iterative Krylov solvers}

One option comes again from iterative refinement \cite[sec~4.1]{ecp_mxp_advances_2021} - use a low-precision GMRES as the inner solver for iterative refinement.
For a highly nonsymmetric matrix coming from convection-dominated fluid flow (`BentPipe2D1500'), the convergence rate for FP32 GMRES(50) with FP64 iterative refinement remains similar to that of FP64 GMRES(50), both without preconditioning. A 1.32$\times$ speedup is observed on an NVIDIA V100 GPU. For other problems, the speedups range from 1.24 to 1.36$\times$ \cite{loe_mxp_gmres_2021}.
With polynomial-preconditioned GMRES, the convergence rate for the two setups of the BentPipe2D1500 problem is identical, and the GMRES(50)-IR approach nets a 1.58$\times$ over GMRES(50) in double.
Loe et al. \cite{loe_mxp_gmres_2021} showed that GMRES-IR is faster than a sequential application of single-precision GMRES followed by switch at any iteration count to double-precision GMRES, at least for two different sparse matrices.

The idea of decoupling the precision used for computation in the arithmetic units of compute cores from that used for storing in device global memory has been explored for accelerating the Generalized Minimum Residual (GMRES) solver by Aliaga et al. \cite{aliaga_cbgmres}.
According to the authors, the performance of Krylov subspace solvers on all modern hardware is limited by memory and communication bandwidth, and therefore it makes sense to compress the Krylov basis vectors in GMRES during main (global device) memory operations.
They call this `compressed basis GMRES' (CB-GMRES).
One way to achieve lossy compression with minimal performance overhead is using low precision.
This is done by a software entity called `memory accessor'. The accessor loads and stores variables into memory from registers automatically casting to and from the storage precision and arithmetic precision, transparent to the kernel operations.
However, this may lead to some loss of orthogonality and thus some degradation in convergence behaviour of the solver. The hope is that the performance benefits from needing to access less memory (owing to low-precision storage) offsets any increase in the number of iterations required to converge.

In their results \cite{aliaga_cbgmres} on Suitesparse matrices, computing in double precision but storing the basis vectors in FP32 or scaled int32 (fixed-point) generally achieves the same residual accuracy as fully double-precision GMRES.
Storing the basis vectors in FP16 or scaled int16 leads to a somewhat worse residual level, but still does better than using FP32 for everything.
After accounting for any increase in iteration count, CB-GMRES with storage in FP32 yields net speedups over regular FP64 GMRES for most classes of matrices from Suitesparse, though it also yields a slowdown on one of the tested classes. On average over all the tested matrices, it yields a speedup of 1.4$\times$.

There has been work on developing communication-avoiding Krylov subspace methods in order to improve their accuracy using mixed-precision techniques \cite{carson_2021_sstep_mxp}.

Graillat et al. \cite{graillat_adaptive_spmv_2024} developed an adaptive precision algorithm for sparse matrix-vector product based on the magnitudes of the matrix entries and used it to accelerate Krylov subspace methods.
Jang et al. \cite{jang_augmented_gmres_2025} have proposed a mixed precision `augmented' GMRES method.

In this regard, the HPG-MxP benchmark \cite{hpgmp} is quite relevant. Derived from the High Performance Conjugate Gradient (HPCG) benchmark \cite{hpcg}, it uses a GMRES solver for a sparse structured numerically nonsymmetric matrix and is designed to allow the use of mixed precision anywhere in the solver except the residual and solution updates.
Using a state-of-the-art optimized implementation, it has been demonstrated \cite{kashi_hpgmxp_2025} that GMRES-IR using double-single mixed precision achieves a 1.6$\times$ speedup over double precision GMRES on AMD and NVIDIA GPUs from the single node scale all the way to a full exascale supercomputer.

\subsection{Preconditioning}

Preconditioning refers to transforming a linear or nonlinear system of equations such that the solution remains the same, but the transformed system is easier to solve by an iterative method. This is typically a very important consideration in fast and scalable solution of large sparse linear systems of equations arising in many scientific problems.

Flegar et al. introduced \cite{adaptive_blk_jacobi} an adaptive-precision block-Jacobi preconditioner. Given a blocking scheme, typically up to 32 for efficiency on NVIDIA GPUs, the algorithm computes the condition number of each block, and based on the condition number, selects a precision format for each block independently.
The number of significant digits to be preserved in the computed preconditioner is a tunable parameter, typically 1 or 2.
All this requires careful implementation and optimization of data structures and GPU kernels, which the authors carry out.
In experiments on 76 matrices from the Suitesparse collection, this method typically yields between 10\% and 30\% speedup. However, for some problems where none of the blocks are well-conditioned, it may give a slowdown of 50\%.

G\"obel et al. \cite{gobel_mixed_sai_2021} implemented a preconditioned BiCGStab solver using variants of mixed-precision implementations of sparse approximate inverse based preconditioners using the memory accessor introduced in the previous subsection.
All computations are in double-precision. All matrices and vectors, except the sparse approximate inverse values, are stored in double precision. The values of the sparse approximate inverse, using the memory accessor, are stored in lower precision.
They test their solver on many Suitesparse matrices. They found that storing in single precision almost always preserved convergence to within 3\% of iterations.
For nonsymmetric matrices, storing in half precision, however, compromises convergence for half the tested matrices.
For symmetric matrices, half precision is more viable and preserves convergence more often.
After accounting for any extra iterations needed, the speedups obtained are quite low - for some cases, one may obtain a 20\% speedup, but the average speedup is nearly non-existent for nonsymmetric matrices and 8\% for symmetric matrices.

\subsection{Multigrid}

Multigrid methods use a hierarchy of discretized problems (matrices or nonlinear systems) to solve the original problem in an optimally scalable manner. For sparse linear systems of equations, algebraic multigrid (AMG) methods are popular scalable preconditioners.
One notable detail about AMG methods is that they use sparse matrix-matrix products in setting up the hierarchy of coarser problems. For nonlinear problems solved by Newton-like methods, this is a non-trivial cost. With modern matrix units on GPUs, it is of high interest to accelerate this operation using mixed-precision hardware-supported instructions.

Tsai et al. have developed mixed-precision algebraic multigrid capabilities \cite{tsai_mxp_amg_2023, tsai_3precision_mxp_amg_2023, tsai_mxp_amg_2024}.
They test their mixed-precision implementation with double-precision storage and operations on the fine-grid but single-precision storage operations on all the coarser AMG levels \cite{tsai_mxp_amg_2023}. The problems include diffusion with discontinuous coefficient discretized by a finite element method, as well as some Suitesparse matrices. AMG is used as a preconditioner in a conjugate gradient (CG) solver.
Going from the double-precision solver to the mixed-precision one, there is typically a minor increase the required number of solver iterations, but a reduction in time-to-solution owing to the low-precision operations on all levels except the finest.
Speedups range from 1.04$\times$ to 1.1$\times$ depending on the problem and the GPU architecture.

More care needs to be taken to leverage half-precision arithmetic \cite{tsai_3precision_mxp_amg_2023}. A naive use of half-precision multigrid levels in the AMG  hierarchy can be catastrophic - with effects ranging from overall slowdowns due to slow convergence rate to outright divergence.
The reasons for this are threefold: for some matrices, the matrix values are out of the range of FP16, for some, the Jacobi smoother is out of range because the inverses of some diagonal values are out of range, and for others, the residual underflows on a coarser FP16 level, thus contributing zero correction to the upper level and offering no preconditioning benefit. The authors attempt mitigation by scaling the matrices and keeping the vectors on an AMG level at higher precision when the matrix precision is FP16.
Running with the configuration `DP-SP, HP' (double-precision vectors on the finest level, single precision all other levels, and half-precision matrix on all levels), convergence is obtained in most cases with speedups over `DP,DP' between 1.1$\times$ and 1.34$\times$ with V-cycle on an NVIDIA H100 GPU.

\subsection{Discrete Fourier transform}
\label{sec:fft}

Large-scale distributed Fast Fourier Transform (FFT) is dominated by inter-process communication time, reportedly up to 97\% \cite{ecp_mxp_advances_2021}. Thus, there have been efforts to compress the communication volume and thus reduce the time required for it.
One simple approach is to cast the data from FP64 to FP32 before communication. In their analysis, this approach gives an error of less than $10^{-7}$, compared to almost $10^{-6}$ for doing everything in FP32. Thus, similar to the memory accessor, a small gain in accuracy can be achieved compared to only FP32, while the time-to-solution remains the same as that using just FP32.
It may be pointed out that for this use case, compression algorithms such as ZFP may work better than simply casting to a lower precision \cite{ecp_mxp_advances_2021}.
This kind of compression can also be integrated into MPI, potentially yielding performance benefits through pipelining and even hardware acceleration in the future.

Sorna et al.\cite{sorna_mxp_fft_2018} propose a recursive radix-4 FFT that utilizes mixed precision matrix-matrix products (GEMMs) on tensor cores.
Their method utilizes multiword arithmetic (section \ref{sec:multiword}) split the FP32 input vector at the lowest recursion level into two FP16 vectors.
Because the Fourier transform is a linear operator, FFT distributes over the scaled summation of the 2-word split input.
The fact that a radix-4 FFT is used means that the operation at the lowest recursion level is a batch of matrix-vector products with the $4\times 4$ FFT matrix which is binary. Thus, the main operation that uses tensor cores is batched GEMM.
While their implementation does not outperform cuFFT, it is a promising step towards a mixed precision FFT algorithm.

Li et al. presented \cite{li_tcfft_2021} a FFT implementation that uses tensor cores and achieves accuracy comparable to using cuFFT in FP16 precision. It does not utilize other precision formats and thus achieves far lower accuracy than purely FP32 implementations, unlike Sorna et al. \cite{sorna_mxp_fft_2018}.
That being said, if one is satisfied with FP16 accuracy, their implementation (see section \ref{sec:lib:fft}) provides substantial speedups over cuFFT - between $1.1\times$ to $3\times$.

Zhao et al. \cite{zhao_mfft_2023} propose their `MFFT' algorithm for distributed FFT that gains speedups at the cost of some accuracy by computing in high precision but communicating in low precision.

\subsection{Nonlinear solvers}
\label{motifs:nonlinear}

Most real-world problems require the the solution of systems of nonlinear equations.
These do not always boil down to a sequence of linear systems, and even when they do, there are additional considerations for the robustness and performance of the nonlinear problem.
This is further complicated by the issue that the properties of a nonlinear solver are highly dependent not only on the specific physical system it is solving, but also where the simulation currently is in its state space (in the case of non-convex problems).

Several parts of nonlinear PDE solvers, especially in time-implicit methods, are traditionally performed using more or less ad-hoc approximations.  Eg.,
\begin{itemize}
    \item Jacobian matrices in Newton loops are computed using lower order accurate discretizations compared to the residual or right-hand side $\bld{r}(\bld{u}_n)$.
    \begin{equation}
        \frac{\partial\bld{r}}{\partial\bld{u}}(\bld{u}_n) \Delta \bld{u} = -\bld{r}(\bld{u}_n).
        \label{eq:newton}
    \end{equation}
    The Jacobian matrix may also be frozen for a few Newton iterations, only updating the right-hand side.
    \item Preconditioners within linear solvers may be computed from a Jacobian matrix of lower-order discretization.
\end{itemize}
These are candidates for acceleration with lower precision, as was seen earlier with CFD and weather applications.

Kelley has published a paper on the analysis of mixed-precision Newton's method in the recent past \cite{kelley_newton_2022}. He showed that with the Chandrasekhar H-equation, depending on the problem parameter, there is typically no difference in the convergence curve when switching to single precision for the Jacobians and linear solves. However, using half precision can significantly degrade convergence, and lost quadratic convergence when it is achievable by single and double precision.

\subsection{Eigenvalue solvers}
\label{sec:eigs}

Finding the eigenvalues and associated eigenvectors
of a square $N\times N$ matrix, $A$, has computational
cost $O(N^3)$, but is challenging to parallelize.
The traditional method is to apply Arnoldi iteration
(or Lanczos, for Hermitian $A$) to reduce the matrix to Hessenberg
(or tridiagonal) form.  Householder transformations and two-step reductions
have also been used.  At the same time, this builds an orthonormal matrix
-- a Krylov subspace of $A$.
This reduces the problem to diagonalization of the resulting Hessenberg matrix,
which is then solved by implicitly shifted QR, QZ, or pole-swapping methods \cite{poleswap}.
 
A simple way to decrease computation time is to perform
the ($O(N^3)$) reduction in low precision, diagonalize in high precision,
and then refine the resulting solutions to higher precision.
Unfortunately, special care must be taken to separate nearby eigenvalues,
whose eigenvectors cannot be distinguished from each other in low precision.
Tsai, Luszczek and Dongarra \cite{tsai_dsyev_2022} showed the SICE and SICE-SM algorithms
to handle this problem while running the refinement in blocks, leading to high arithmetic intensity.
In the single-step reduction, they showed a speedup of 11\% on NVIDIA V100 GPUs (1:2 double to single FLOP ratio)
from using mixed precision, and 78\% on an NVIDIA GTX1060 consumer graphics card with a 1:32 double to single FLOP ratio.
Again, the distribution of eigenvalues was important.
Using matrices with geometrically distributed eigenvalues weighted toward
the lower limit (e.g. in $[10^{-7},1]$) was shown to require to more iterations
because of the close spacing.

Speedups from mixed precision also compete with other optimizations.
The multi-step reduction has half the overall runtime of the single-step reduction on
CPU, and starts to become dominated by diagonalization because of its
low arithmetic intensity.  In order to finally achieve a 20\% speedup from mixed
precision, they carefully re-ordered the computational steps so that forward and
reverse transformation steps could be carried out on the GPU concurrently with
diagonalization on the CPU.

The ELPA-AEO collaboration\cite{elpa_aeo_2019} showed the results of several experimental
schemes for mixed precision eigenvalue solvers in the context of quantum chemistry.
For this application, only the smallest few ($O(1-10\%)$) of the eigenvalues and eigenvectors
are needed -- motivating subspace iteration methods.  The BEAST-P subspace refinement iteration applies
a polynomial function of the matrix to a set of vectors (current subspace) and then diagonalizes
the result.  The authors showed that the convergence error of subspace
iteration in single precision is just as fast as double precision until the
residual decreases to around $10^{-6}$.  At this point, switching to double precision
computations continues to decrease the residual.  For other steps in quantum chemistry
(computing $A X$ and $X^T X$, where $X$ is the subspace), increasing the precision to FP128
{\em decreased} the overall runtime since these steps have lower dimension but strongly
impact the number of steps needed to converge.

Kodali, Ramakrishnan, and Motamarri demonstrated a general mixed-precision eigenvalue solver for sparse or dense matrices based on Chebyshev filtered subspace iteration (ChFSI) \cite{kodali_2025}.
They separated out the residual at each matrix multiplication step in the Chebyshev recurrence.
The result (R-ChFSI) effectively suppresses errors made in matrix-vector products.  This allows the use of reduced precision matrix-vector products, and (in the case of the generalized eigenvalue problems) approximate inverses.
Using standard ChFSI, reduced precision presents a severe accuracy limit (about $10^{-4}$ instead of $10^{-12}$ achievable with FP64).
Tests were done on problems of dimension up to 7.6 million to filter eigenspaces of dimension 14 thousand. They demonstrated that R-ChFSI converges in a similar number of steps and achieves FP64 accuracy as in standard ChFSI.
On the NVIDIA H100 GPU, they achieved speedups over FP64 of 1.7 for FP32 and 2.0 for TF32.

These examples show that iterative methods have many potential
avenues to pursue speedup from mixing lower precision and higher
precision computations.
The situation is similar to multigrid and nonlinear solvers.
Subspace iteration methods are promising
because a subspace found in low precision has numerical errors
that can be systematically corrected by adding high-precision information.

\subsection{Ordinary differential equations}
\label{sec:ode}
Systems of ordinary differential equations of the form
\begin{equation}
    \frac{d\bld{y}}{dt} = \bld{f}(\bld{y}),
\end{equation}
$\bld{y}:[0,T]\rightarrow \mathbb{R}^n$, $\bld{f}:\mathbb{R}^n \rightarrow\mathbb{R}^n$ arise in many scientific problems, typically as the result of spatial semi-discretization of a time-dependent problem.

Recently, there have even been studies of how to use lower precision while solving systems of ordinary differential equations.
Grant \cite{grant_perturbed_rk_2022} proposed mixed-precision fully implicit additive Runge-Kutta (MP ARK) methods.
He used B-series analysis \cite{bseries} to split the discretization error into the scheme error and the perturbation error, where the latter comes from using lower precision to evaluate the right-hand side.
Using this analysis, he was able to explain that a mixed precision implementation of the implicit midpoint method loses accuracy (from 2nd order to 1st) because of high perturbation error.
He then shows that adding additional explicit steps in high precision solves the issue, and gives an accurate solution even with a precision of just 1 digit for the low precision implicit stage!
Because the implicit stage typically costs a lot more than an explicit stage especially for nonlinear ODEs, this can result in significant speedup.
The analysis is then also used to derive novel third-order accurate SDIRK (Singly Diagonally Implicit Runge Kutta) schemes which retain their accuracy in a mixed precision setting using additional (comparatively inexpensive) explicit stages.
The caveat is, however, that a stability analysis is omitted.
It is thus difficult to say whether the results would hold for most complex problems such as turbulent fluid flow and plasma physics.

Burnett et al. \cite{burnett_ark_stability_eval_2022} tested the real-world stability and performance of the mixed precision ARK methods introduced by Grant, on the 1-dimensional viscous Burgers equation.
While this equation captures some properties of fluid flows, it is quite simplified, especially because only one spatial dimension is considered. Therefore, these results should be taken with a pinch of salt.
They obtained their results on modern single CPUs from IBM and Intel, but their results are likely still relevant since the problems are most likely limited by memory bandwidth.
One must also note that because different time integration schemes have different orders of accuracy with respect to the time step, it is possible for different methods to be advantageous at different error levels.
With two explicit correction steps, Burnett et al. find that the implicit midpoint method (having convergence order 2) in double-half mixed precision is about 2-4$\times$ faster than the fully double precision implementation.
Their mixed-precision formulation of a 3rd order singly diagonal implicit Runge-Kutta (SDIRK) method delivers up to roughly an 8$\times$ speedup over the fully double precision implementation!
These results hold for error levels roughly between $10^{-10}$ and $10^{-13}$. Beyond $10^{-13}$, the errors of all methods up to 64-bit precision start increasing after the time step becomes too small.

Dravins et al. \cite{dravins_rk_2024} implement and test Grant's MP ARK schemes on three-dimensional linear PDE problems.
3D heat equation is tested at error levels of $10^{-3}$ to $10^{-8}$ while the linear advection equation is tested at error levels of $10^{-2}$ to $10^{-5}$. With the linear solve at each time step performed by low precision preconditioned Krylov solvers, they find that no significant loss of accuracy.
On the other hand, because they use a highly tailored tensor product preconditioner which is optimal for the heat equation on Cartesian grids, the use of low (single) precision for the Conjugate Gradient solver leads to slower convergence, roughly doubling the number of iterations required.
Since the preconditioner is not optimal for the advection equation, its GMRES linear solver suffers less from low precision.
Finally, they measure the performance of all required linear solves in a simulation using the mixed precision 4s3pC integrator.
At moderate time steps and practical solver tolerances, they observe a speedup of 1.2$\times$ for the heat equation and about 1.7$\times$ for the advection equation.
This difference is primarily because of the weaker convergence rate of the low precision linear solver in the case of the heat equation. This is an artefact of the highly customized preconditioner; in more general-purpose usage, one would use off-the-shelf preconditioners like parallel incomplete LU factorization. In this case we believe the speedups would be closer to those seen by Dravins et al. for the advection equation.
We note the caveat that the overall application speedup, which would include assembly of the residuals in low and high precision, is not addressed by the study.

Croci and de Souza investigated \cite{croci_explicit_rk_2022} explicit stabilized Runge-Kutta methods. In such methods, there is no nonlinear or linear system to be solved in each stage. However, compared to typical explicit Runge-Kutta methods, they perform additional stages to increase the stability bounds of the scheme.
This is advantageous if the increase in the allowed time step size allows one to reduce the number of time steps required to the extent that it compensates for the cost of the additional stages per step and results in savings.
Since the stability bound grows quadratically with the number of stages, this is observed in many cases.
Croci and de Souza derive mixed precision methods that retain first or second order as the time step $\Delta t \rightarrow 0$.
\begin{greenbox}
The method of Croci and de Souza \cite{croci_explicit_rk_2022} hinges on a careful construction of a low-precision approximation of the secant $\hat{\Delta}\bld{f}_j \approx \Delta\bld{f}_j = \bld{f}(\hat{\bld{y}} + \hat{\bld{d}}_j) - \bld{f}(\hat{\bld{y}})$ in each stage, such that
\begin{align}
    \hat{\Delta}\bld{f}_j &= \Delta\bld{f} + \mathcal{O}(\epsilon\Delta t) \quad \forall j,\, \text{or} \\
    \hat{\Delta}\bld{f}_j &= \Delta\bld{f} + \mathcal{O}(\Delta t^2) \quad \forall j
\end{align}
for a small constant $\epsilon$.
\end{greenbox}
In general, their method requires the Jacobian-vector product, which is significant disadvantage for an explicit scheme.
For discretized nonlinear diffusion problems, they demonstrate that their methods, using bfloat16 as the low precision, retain 1st or 2nd order of accuracy as $\Delta t \rightarrow 0$, as opposed to a na\"ive mixed-precision implementation. In all their experiments, they scale the matrices by the maximum entry.
However, they do not report on savings in time-to-solution not energy, since they emulate bfloat16 on CPUs.

Balos et al. \cite{balos_exp_odes_2023} derive a reformulation of the second-order Rosenbrock-Euler method that retains its accuracy even while using single precision. They demonstrate its effectiveness on a 2D advection-diffusion-reaction PDE problem.
While they do not provide performance numbers, this work adds to the mounting evidence that research and development of mixed precision algorithms for a variety of time integrators could yield benefits.

%\subsection{Use of tensor cores}
%Acceleration of coarse level generation in Algebraic Multigrid (AMG) methods

\subsection*{Common observations}

We note some cross-cutting issues across different mixed-precision motifs.
\begin{itemize}
\item In the case of iterative refinement (IR) methods, somewhat ill-conditioned problems require many iterations to converge. This may, in some cases, negate the advantage coming from the use of low-precision factorization.
\item There is often the necessity to create copies in different precision formats. In some cases, this may severely affect overall application performance.
One remedy that may be applied sometimes, though it comes with its downsides, it `in-place' casting and conversion of data to lower precision \cite{walden_unstructured_mixed_2019}, \cite{amestoy_sparse_ir_2023}.
\item For methods tested on sparse matrices, the practically attained speedup and accuracy can depend significantly on the source of the matrix in terms of science domain and physical parameters.
\end{itemize}

\begin{table}
    \centering
    \begin{tabular}{|c|c|c|c|}
    \hline
    \textbf{Method} &  \textbf{Motif} & \textbf{MxP alg. class}  & \textbf{Possible speedup} \\
    \hline
    Ozaki I   &  \emph{DMM} &  split  &  $0.5\times$ \\
    Ozaki II  &  \emph{DMM} & split   &  $1.23\times$ \\
    Adaptive SpMV & \emph{SpMV} & split & $1\times$ -- $4\times$ \\
    LU-IR     &  \emph{DF}  &  MxP-R &  3$\times$ - 9.5$\times$      \\
    LU-IR     &  \emph{SpF} & MxP-R &  $\sim 1.6\times$    \\  % Or slowdown of 0.57$\times$ 
    GMRES-IR  &  \emph{SpItLA} & MxP-R & $\sim 1.6\times$  \\ % or slowdown of 0.57$\times$
    CB-GMRES  &  \emph{SpItLA} & split & $\sim 1.4\times$ \\
    Adaptive blk. Jacobi & \emph{SpItLA} & MxP-inL & $1.1\times$ -- $1.3\times$ \\
    SpAI, ISAI & \emph{SpItLA} & MxP-inL &  1$\times$ -- 1.2$\times$ \\
    Multigrid &  \emph{SpItLA} & MxP-inL &  1.1$\times$ -- 1.34 $\times$ \\
    SICE-SM &  \emph{Eigs}     &  MxP-R & $1.2\times$ (real) -- $1.45\times$ (complex) \\
    BEAST-P &  \emph{Eigs}     &  MxP-R & $1.2\times$ \\
    R-ChFSI &  \emph{Eigs}     &  MxP-R & $2.0\times$ \\
    MP-ARK  &  \emph{ODE}      & MxP-R & 1.7$\times$ \\
    \hline
    \end{tabular}
    \caption{Possible speedups from different mixed-precision algorithms. \emph{DMM} refers to dense matrix multiplication, \emph{SpMV} refers to sparse matrix vector product, \emph{DF} refers to dense factorization, \emph{SpF} refers to sparse factorizations,  \emph{SpItLA} refers to sparse iterative linear algebra, \emph{Eigs} refers to eigenvalue problems and \emph{ODE} refers to ordinary differential equations. Classes of mixed precision algorithms (introduced in the text) are `MxP-R': mixed precision with refinement, `MxP-inL': inner low precision, `split': splitting methods.}
    \label{tab:methods}
\end{table}
In table \ref{tab:methods}, we summarize the range of speedups that developers of mixed-precision algorithms have observed. 

%\begin{table}
%    \centering
%    \begin{tabular}{|c|m{0.8in}|m{0.8in}|m{0.8in}|m{0.8in}|}
%    \hline
%    Appl. domain &  Dense factorization & Sparse factorization & Iterative methods &  Fourier transforms \\
%    \hline
%    CFD - aero, astro &    \cellcolor{red!50} 0      & \cellcolor{green!25} 6 & \cellcolor{green!55} 8 & \cellcolor{red!40} 1 \\
%    CFD - turbulence research & \cellcolor{red!50} 0 & \cellcolor{red!50} 0 & \cellcolor{red!40} 1 & \cellcolor{green!50} 9 \\
%    Weather   & \cellcolor{red!50} 0       & \cellcolor{red!20} 3   & \cellcolor{green!40} 7 & \cellcolor{red!40} 1 \\
%    Quantum chemistry &\cellcolor{green!55} 8 & \cellcolor{red!40} 1 & \cellcolor{red!40} 1 &  \cellcolor{red!40} 1 \\
%    Sequence similarity & \cellcolor{green!65} 9 & \cellcolor{red!50} 0  &  \cellcolor{red!50} 0 &  \cellcolor{red!50} 0 \\
%    \hline
%    \end{tabular}
%    \caption{Relevance, on a scale of 10, of mixed-precision methods to different application areas}
%    \label{tab:appl_methods_relevance}
%\end{table}

Broadly, the mixed precision algorithms can be divided into three classes.
\begin{enumerate}
\item Low precision: This does not really refer to mixed precision algorithms, but rather to computations that can be done in reduced precision without significantly affecting the accuracy of the result, as in some turbulence research and weather prediction applications.
\item Inner low precision (MxP-inL): Numerical methods where some `inner' operations are inherently rough approximations, such as preconditioners, multigrid smoothers etc., that can thus be performed in low precision. One may need to be careful to avoid numerical instability leading to NaNs or Infs, though the solution is good if the algorithm successfully completes.
\item Mixed precision with Refinement (MxP-R): Algorithms where iterations or additional steps with some operations in high precision are used recover the accuracy of low precision computations. Examples include iterative refinement and explicit corrector steps in mixed precision implicit Runge-Kutta time steppers.
\item Splitting schemes (split): Algorithms to directly approximate some classes of high precision operations using split low precision numerical representations, such as the multiword, Ozaki and adaptive SpMV \cite{graillat_adaptive_spmv_2024} schemes.
These are fundamentally based on the idea that numbers with relatively small magnitude can be represented in smaller formats without catastrophic rounding errors. Here, we also include methods that selectively use high precision, for example, to accumulate long sums.
\end{enumerate}
\section{Impact of mixed-precision techniques on resource utilization}

As computational scientists, we wish to minimize the resources our applications utilize for a given outcome. For our purposes here, these are time on a computer, memory footprint and energy consumed by the computer while running our application.
So far, we have mainly focused on the time-to-solution improvements offered by various mixed-precision algorithms in various applications and motifs.
We now briefly turn our attention to memory footprint and energy consumption.

New GPU architectures are increasingly targeted at AI workloads. Perhaps it should come as no surprise that they excel at this targeted workload in terms of energy efficiency as well.
We have discussed earlier in this article, eg. while discussing the work of Chakravarty \cite{speech_ml_enegy_2024}  and Kermani et al. \cite{transformer_energy_efficiency_2025}, how low precision formats like FP8 and FP16 typically need much less energy than single precision for AI workloads.
%\begin{figure}[h]
%    \centering
%    \includegraphics[width=0.98\linewidth]{fig/nvidia-computex-gpu-performance-energy-over-time.jpg}
%    \caption{GPU performance and energy consumption for large language model training over the years; courtesy NVIDIA}
%    \label{fig:energy_llm}
%\end{figure}
The question is: how much of this increased efficiency applies to computational science workloads that are not AI?

For workloads that depend on dense matrix multiplication, the energy efficiency impact of splitting techniques have been recently demonstrated.
Ootomo et al. \cite{ootomo2024dgemm} found that the INT8-based Ozaki scheme consumes about the same amount of power as DGEMM on the NVIDIA A100 GPU. Thus, the energy efficiency is directly proportional to the time to solution, and the Ozaki scheme with INT8 attains less than half the GFLOPS/Watt that DGEMM attains.
On the latest Blackwell generation, NVIDIA has reported [private communication] 4-34$\times$ increase in performance per watt for the GEMM part of a few scientific workloads have been observed on recent hardware.
However, as explained in that section, Ozaki splitting also incurs a higher memory footprint to the tune of $2.5\times$.

How about workloads that do not depend on large dense matrix multiplications?
There have been efforts to study energy savings from mixing lower precision formats in scientific computing going back several years. In 2010 on multicore CPUs, Anzt et al. \cite{anzt_cfd_energy_2010} approximately estimated a 50\% energy savings using mixed-precision IR and GMRES to solve some sparse linear systems from a two-dimensional CFD problem.
Haidar et al. \cite{azzam_half_dense_energy_2018} carried out dense linear system solves on Intel Xeon Phy Knights Landing and NVIDIA V100 GPUs to find 50\% and 80\% energy reduction respectively by using a FP16-FP32-FP64 mixed precision IR method.

We noted in section \ref{sec:appl_conclusions} that going from double to single precision can provide significant reductions (34-38\%) in energy consumption for memory bandwidth-limited scientific workloads such as earthquake simulations.
This reflects the fact that memory fetches and stores cost a significant amount of energy, and reducing the data movement volume from reduces the energy consumption.

\section{Libraries with mixed-precision capabilities}

Several research groups have made their mixed precision capabilities available in software libraries. In particular, the ECP mixed-precision effort has led to the inclusion of well-tested mixed-precision algorithms in open-source libraries new and old.
Abdelfattah et al. provide \cite{icl_mxp_survey_2020} a concise overview, current in 2020, of existing library support for mixed-precision numerical methods.
In table \ref{tab:appl_methods_libraries}, we provide an overview of the motifs available in the different libraries surveyed and the extent of mixed-precision support.

We admit that our focus on certain application domains (section \ref{sec:appls}) has led to the omission of some libraries that are otherwise of interest. An example is sparse direct solvers that are critical in several fields including electricity grid simulations.
As such, we do not claim that this list of libraries is close to comprehensive; it serves instead as a starting point comprising those libraries that either have high support for mixed precision algorithms or are particularly relevant to the applications of interest to us.

We also omit vendor-specific libraries such as those in NVIDIA's CUDA toolkit. Though some of these libraries provide useful mixed precision algorithms or building blocks, we focus here on open-source and cross-platform libraries that can be used on a wide variety of hardware platforms.

\subsection{Magma}

Magma \cite{magma_2010} is a library containing dense (and sparse) linear algebra routines, including those in BLAS and LAPACK. It targets single-node high performance on GPUs and other `many-core' architectures.
It has a CUDA version, an OpenCL version and an OpenMP version, though the CUDA version is the primary one.
The GPU kernels are optimized using auto-tuning methods including code generation, and loop transformations.
In addition to routines for factorization and linear solvers, it also has highly optimized kernels for eigenvalue and singular value computations \cite{magma_hessenberg_2010}.
Magma-sparse contains many fine-grain parallel (GPU-friendly) algorithms, such as iterative incomplete LU factorization \cite{chow_parilu_gpu_2015}, asynchronous iterations etc.

Magma has implementations of several variants of mixed-precision iterative refinement \cite{magma_half_2017}, including GMRES-IR \cite{magma_mxp_2018}.
It even contains iterative refinement for least-squares problems using QR factorization.

\subsection{Slate}

SLATE \cite{slate_2019} provides a cross-platform GPU-enabled distributed dense linear algebra and solver library written in modern C++.
It provides substantial infrastructure for mixed-precision developments by templating all functionality on a generic scalar type, as well as providing overloads of functions that apply to all numeric scalar types.
This enables one to both write numerical code in a precision-agnostic manner, and also to mix and match different precision formats for different operations.
It provides templated C++ wrappers for BLAS subroutines to aid this effort. The functionality provided by these wrappers is aware of streams or queues on each of the targeted GPU architectures \cite{slate_2024}.
SLATE also contains a mixed-precision iterative refinement implementation.

In addition to being flexible on the scalar type, its main features are the use of arbitrary distribution of matrix blocks (not just block-cyclic, even though it is optimized for this) and task-based parallelism using OpenMP tasking constructs.
SLATE maintains memory consistency across different available memory spaces on a tile-by-tile basis. Tiles are `mirrored' (copied) wherever needed and deleted when no longer needed. Initial copies of tiles are labelled as the `origin', remote copies are made when needed and deleted, and it is ensured that the origin is updated.

\subsection{Ginkgo}
Ginkgo \cite{ginkgo-toms-2022} is a modern C++ library for mixed-precision sparse linear algebra computations.
It supports multicore CPUs, NVIDIA GPUs, AMD GPUs and Intel GPUs through the use of a backend architecture that allows a common high-level implementation of solvers while accommodating separate highly optimized kernel implementations for each the different operations required for each backend.
Ginkgo follows a rigorous unit-testing, integration-testing and benchmarking workflow in order to ensure correct and performant software.
It has integrations with several science libraries and applications, including DEAL.II \cite{dealII95}, MFEM \cite{mfem}, SUNDIALS \cite{sundials_2022} and the XGC \cite{xgc_2017} plasma simulation code.
It supports mixed- and multi-precision computation in two ways. Firstly, all functionality, including, solvers, preconditioners, matrix formats etc. are templated on the scalar type. Thus, users can build mixed-precision applications using the building blocks provided by Ginkgo. Secondly, there is explicit support for some mixed precision algorithms such as mixed-precision block Jacobi \cite{adaptive_blk_jacobi} and mixed-precision algebraic multigrid \cite{tsai_mxp_amg_2024}.
The multigrid support is very modular and flexible, allowing one to choose the precision of the different components at each multigrid level, owing to the library's in-built precision conversion mechanism. The precision conversion of the residual and correction vectors happens `on-the-fly' in the restriction and prolongation operations.

\subsection{Trilinos}
Trilinos \cite{trilinos-website} is a collection of C++-based scientific computing libraries released primarily by Sandia National Laboratory since 2003. The packages mainly focus on performance-portable linear solvers, nonlinear solvers, and numerical optimization. It is relied upon by a large number of other scientific libraries and applications.
Trilinos contains the Belos and Amesos2 packages for sparse iterative and direct solvers respectively \cite{belosamesos2}.
While Amesos2 is an extensible and efficient wrapper for accessing many different sparse direct solvers via a common interface, Belos is a platform-agnostic extensible iterative solver library with modular implementations of many solvers like GMRES, BiCGStab (Bi-conjugate gradient method), TFQMR (Transpose-Free Quasi Minimal Residual method) etc.
Both these libraries utilize modern C++, and can deal with matrices of arbitrary scalar type, thus enabling mixed-precision algorithms.
An example is an adaptive-precision LSQR iterative algorithm in Belos \cite{belosamesos2}.
However, Belos does not provide the flexibility of using different scalar types for the different internal components within one linear solver \cite{loe_mxp_gmres_2021}.

At a lower level of abstraction, these solver libraries are implemented on top of TPetra and Kokkos-Kernels. TPetra is a platform-agnostic templated distributed sparse matrix library \cite{tpetra_2012}. The authors in 2012 had envisaged mixed-precision algorithms and implemented support for arbitrary precision using C++ templates in distributed sparse matrix operations.
Kokkos-Kernels \cite{kokkoskernels_2021}is a sophisticated cross-platform library for sparse linear algebra and graph algorithms, including sparse matrix sparse matrix products (SpGEMM), sparse triangular solves (SpTRSV), sparse matrix addition (SpAdd or SpGEAM) and others.
Similar to the other modern Trilinos libraries, Kokkos Kernels functions and classes are templated on the scalar type, thus enabling the development of mixed-precision algorithms.
All these are implemented on top of Kokkos \cite{kokkos3_2022}, which is the key to the performance portability of not only modern-day Trilinos but also many other scientific applications around the world \cite{kokkos_eco}.

While Trilinos libraries like Belos do not themselves contain mixed-precision algorithm implementations, as promised, researchers have leveraged them to implement solvers like GMRES-IR \cite{loe_mxp_gmres_2021}. A mixed-precision $s$-step conjugate gradient solver was implemented \cite{yamazaki_sstepCG_2022} on top of Kokkos and Kokkos-Kernels.

\subsection{PETSc, Hypre and SUNDIALS}
\label{sec:other_libraries}

PETSc \cite{petsc-efficient} is a longstanding library with varied functionality for distributed sparse problems, including primarily linear solvers and preconditioners, but also nonlinear solvers, time-stepping schemes and numerical optimization methods. It is used extensively by science and engineering applications worldwide.
As of writing, PETSc does not support mixed-precision algorithms in its own code. There are, however, plans to support some level of coarse-grained support for mixed-precision as of 2021 \cite{petsc_portable_2021}.
While discussing PETSc in this article, we also include SLEPc \cite{slepc_2005}, a library for sparse distributed eigenvalue and singular value decomposition (SVD) problems \cite{roman_slepc_2023} built on top of PETSc. Similar to PETSc, it now supports GPUs but not mixed precision methods.

Hypre \cite{hypre} is a library that implements scalable multi-level preconditioners and solvers for grid-based and grid-agnostic simulations. It has been under development since 1993, primarily at Lawrence Livermore National Laboratory.
Hypre, like PETSc, is built for one precision. As of 2021, there was ongoing work to build all precision versions of the important `multi-precision functions' in Hypre so as to allow mixed-precision algorithms \cite{ecp_mxp_advances_2021}.

SUNDIALS \cite{hindmarsh_sundials_2005} is a suite of time integrators and nonlinear system solvers aimed at solving systems of ordinary differentia equations (ODEs) of various types. Like PETSc and Hypre, it is a longstanding open-source software project of the US Department of Energy and was part of the Exascale Computing Project.
It is relied upon by various PDE-based simulation projects.
In recent years, it has added substantial support for GPUs \cite{sundials_2022}. However, similar to PETSc and Hypre, one needs to provide the scalar type during build time, and it supports only one floating point format in the build.
As of writing, the authors are not aware of any mixed precision capabilities in the library, though some mixed precision use cases may be enabled by recent integrations with other libraries such as Ginkgo.

\subsection{Fast Fourier Transform libraries}
\label{sec:lib:fft}
HeFFTe \cite{heffte_2020} is a modern cross-platform distributed discrete Fourier transform (DFT) library.
Because it uses template parameters for the input and output vector types, it supports different scalar types.
The code has been used to investigate compressing communications on-the-fly using reduced precision \cite{heffte_mxp_2022}.

tcFFT \cite{tcfft} is a relatively recent FFT implementation that runs on NVIDIA GPUs and utilizes the FP16 tensor core approach of Li et al. \cite{li_tcfft_2021} to gain $1.1\times$ to $3\times$ speedup over cuFFT.
These are enabled not only by the use of tensor cores, but carefully designed efficient GPU kernels for other tasks required by FFT that do not use tensor cores, such as element-wise multiplication.
However, the algorithm is designed to achieve only FP16 level of accuracy - for small inputs, while FP32 FFT can achieve a relative error level on the order of $10^{-6}$, tcFFT (and FP16 cuFFT) achieve only about 0.01.
As of writing, tcFFT is not actively developed, though the code is available.

\begin{table*}
    \centering
    \begin{tabular}{|c|p{0.7in}|p{0.7in}|p{0.7in}|p{0.7in}|p{0.7in}|p{0.7in}|}
    \hline
    \multirow{2}{*}{Library} & Dense & Sparse & Iterative  &  Pre-  & \multirow{2}{*}{Multigrid} & Fourier   \\
    &factorization& factorization&methods&conditioners& &transform\\
    \hline
    \hline
    Magma & \supportss{4}{0}{0} & \supportss{0}{0}{1} & \supportss{0}{0}{2} & \supportss{0}{0}{1} &  & \\
    \hline
    Slate   & \supportss{3}{0}{2} &            &            &            &  &   \\
    \hline
    Ginkgo  & \supportss{1}{1}{0} & \supportss{1}{0}{1} & \supportss{4}{0}{0} & \supportss{4}{0}{1} & \supportss{2}{0}{1} & \supportss{0}{1}{0} \\
    \hline
    Trilinos &  \supportss{1}{1}{0} & \supportss{2}{0}{1} & \supportss{4}{0}{1} & \supportss{4}{0}{1} & \supportss{1}{0}{4} &  \\
    \hline
    HeFFTe   &       &                 &                 &          &      &  \supportss{3}{0}{2}  \\
    \hline
    \end{tabular}
    \caption{Availability of mixed-precision methods in different libraries. Filled bubbles indicate availability of implementations for the respective motif including mixed-precision support, while empty bubbles indicate availability of fixed-precision implementations only.}
    \label{tab:appl_methods_libraries}
\end{table*}

\section{Conclusions}

\begin{figure}[h!]
\centering
\includegraphics[width=0.98\linewidth]{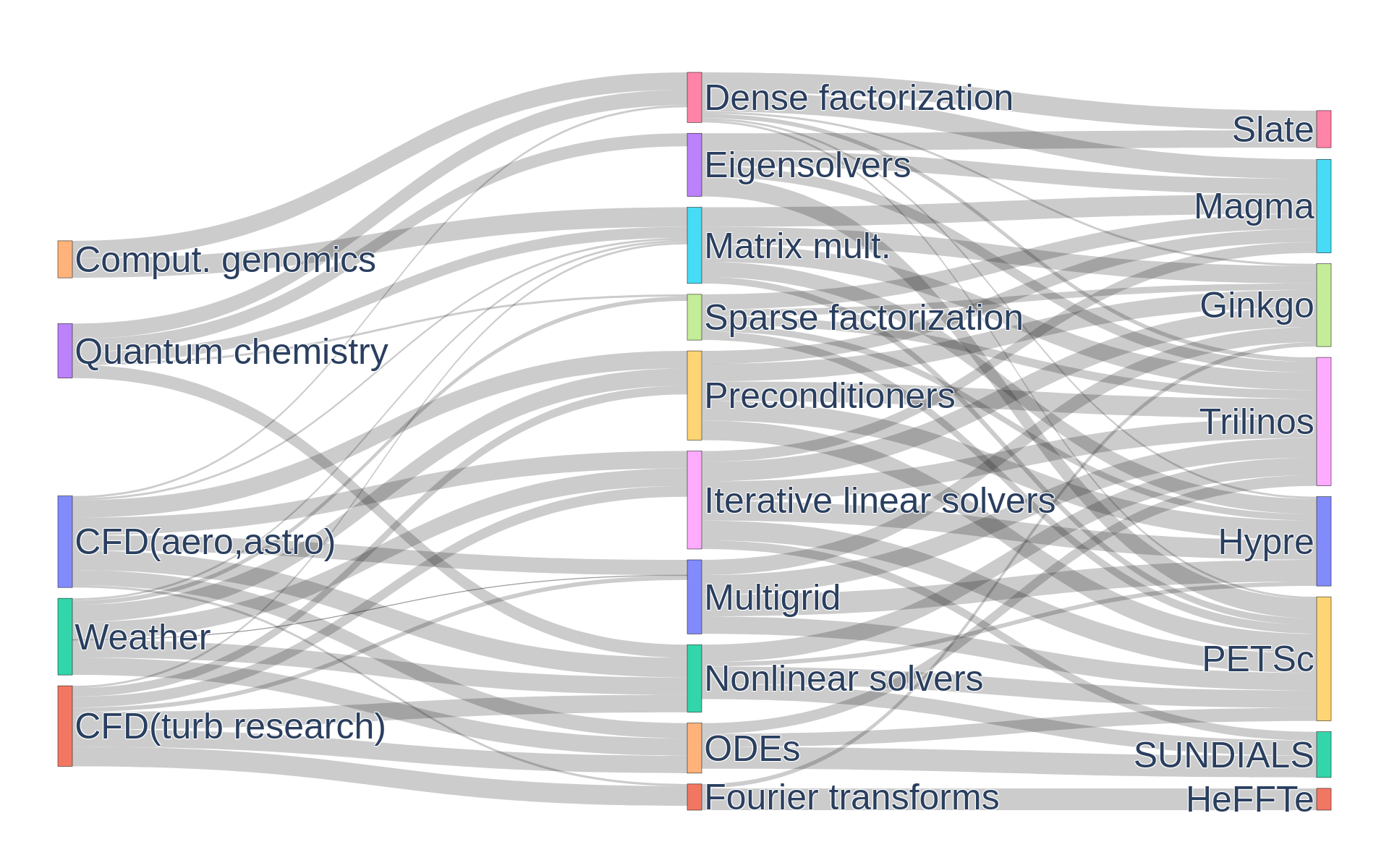}
\caption{Visualization of how much an application domain (left column) depends on different computational motifs (middle column), and what libraries (third column) provide these motifs in fixed (double) precision.}
\label{fig:sankey_fixed}
\end{figure}

\begin{figure}[h!]
\centering
\includegraphics[width=0.98\linewidth]{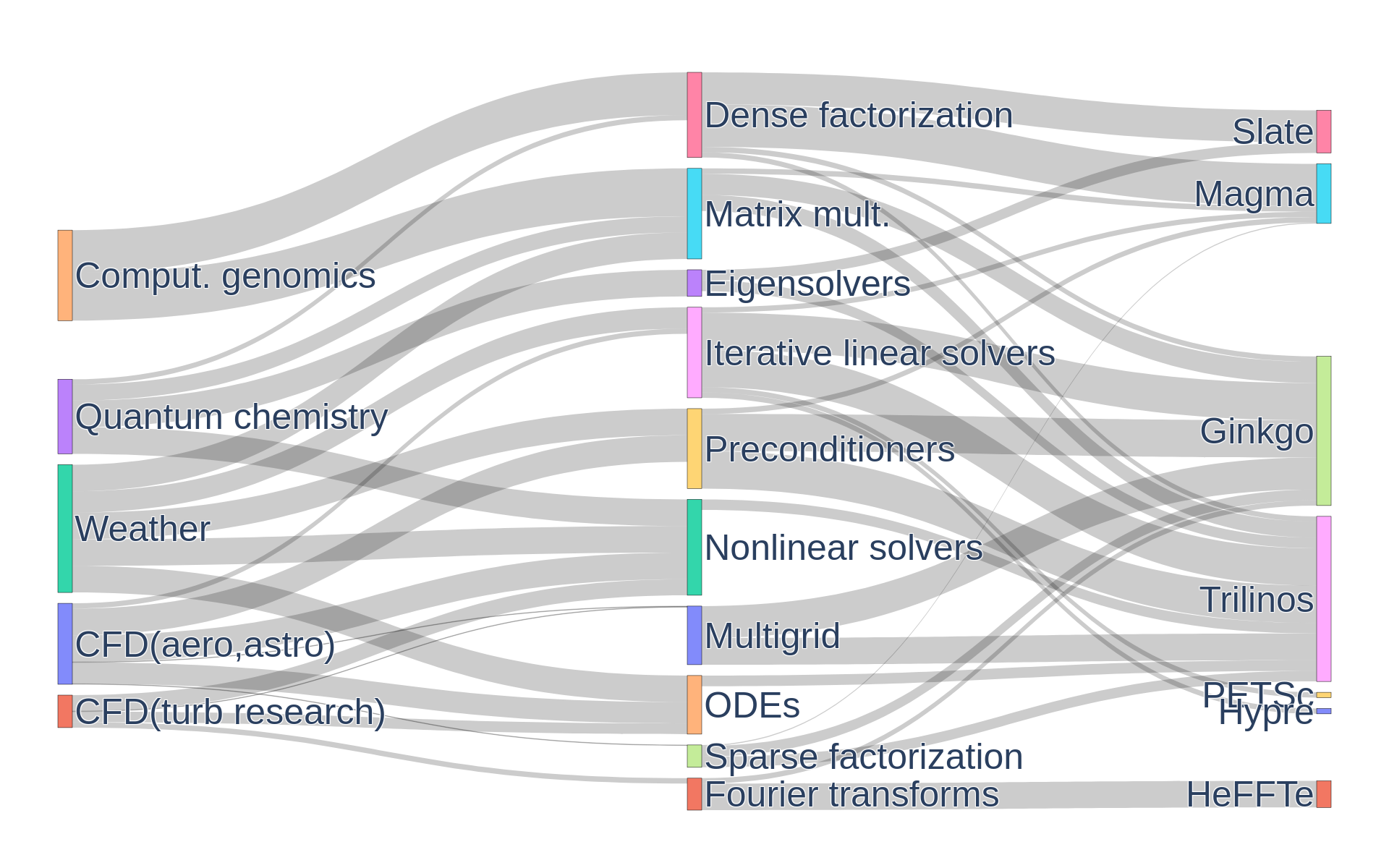}
\caption{Visualization of how much an application domain (left column) makes use of mixed-precision techniques developed for different computational motifs (middle column), and what libraries (third column) provide mixed-precision algorithms for these motifs.}
\label{fig:sankey_mxp}
\end{figure}

Figure \ref{fig:sankey_fixed} shows a subjective visualization of the simulation ecosystem in terms of our surveyed application domains, their usage of different computational motifs, and the libraries that implement these motifs with traditional fixed-precision algorithms.
Figure~\ref{fig:sankey_mxp} shows the usage of mixed-precision algorithms for computational motifs by application areas, along with the libraries providing mixed-precision algorithms for these motifs.
Note that the scale of thickness of links in the two plots is different.

These plots visually express the potential for improvement in mixed precision scientific computing - one can see how links relatively thin out going from fixed double precision to mixed precision, both in terms of applications making use of mixed precision algorithms for different motifs as well as libraries providing mixed precision implementations for different motifs.

Some observations emerge from our review. It is clear that even though CFD applications depend on iterative solvers, they are not using innovations in mixed precision iterative solvers.
This may be partially due to the historical tendency of CFD development teams to develop iterative solver software in-house. More co-design and collaboration would help apply mixed precision innovations by numerical linear algebra researchers to CFD codes.
Another impediment is that established libraries like Hypre and PETSc have been slow to support mixed-precision implementations.

Similarly, the quantum chemistry domain does not fit standard computational motifs provided by existing libraries.
There are not many clear domain-agnostic patterns in quantum chemistry applications which makes it difficult to design high-level libraries to express mixed precision algorithms for quantum chemistry.
For example, the mixed-precision kernels used by DFT-FE fall under distributed dense linear algebra.
However, the distributed dense linear algebra library Slate does not
really provide those algorithms, though it does make efficient and usable implementations easier.
Meanwhile, quantum chemistry applications have not widely adopted useful mixed-precision
libraries for dense linear solvers -- for example in solving the Poisson problem.
Indeed, when they are not performance-limiting, DFT-FE performs these operations on CPUs.
Progress in the field can be accelerated by making mixed precision eigensolver algorithms,
like the ones reviewed in section \ref{sec:eigs}, in cross-platform libraries such as SLATE and SLEPc.

On the positive side, the computational genomics applications we studied have already adopted mixed precision computations to achieve significant speedups while retaining the needed accuracy.
Support for the types of dense matrix multiplication and factorization encountered in genomics applications in readily-accessible libraries can further accelerate the development of more capable techniques and their application to novel datasets.

\section{Perspectives and recommendations}

Dongarra et al. \cite{dongarra_hardware_trends_2024} envisage a scenario in the future in which precision is dynamically adjusted to optimize performance and energy efficiency. We share this vision and believe such a system will need to take into account specific details of different domains and applications, while also leveraging algorithms for different computational motifs.
Recent innovations in achieving high-accuracy matrix multiplication using splitting schemes are a natural fit in this vision, though there remain technical challenges to be overcome in schemes such as those from Ozaki et al.

For some applications, the main effort required is domain-specific error analysis and equation transformations. In these cases, libraries can, at best, provide secondary assistance. This can be seen in the domain-specific ways some weather and climate as well as turbulence research applications tend to use mixed-precision numerics.
Such efforts can be significantly helped by easy-to-use tools to analyze the effect of low precision on the accuracy of simulations and performance. Tools such as RAPTOR \cite{raptor_riken_2025} and SherLogs.jl \cite{kloewer_shallowwater_fp16_2022} are good examples.

Nevertheless, there are low-hanging fruits in terms of utilizing mixed-precision methods for applications that rely primarily on linear algebra routines. These include some applications in quantum chemistry and physics, CFD, and weather and climate.
Iterative refinement with low-precision inner solvers is a relatively simple way of exploiting low precision hardware units. In other words, one may try, with relative ease, to use a pre-existing effective FP64 solver as an inner solver for IR by uniformly changing the number format.
This is especially relevant for applications that depend on dense linear algebra operations, such as quantum chemistry and sequence similarity, which can leverage tensor cores and matrix processing units.

For applications that are limited by memory bandwidth, there is still substantial algorithm and software technology available. However, the maximum realizable speedup on current memory technologies is limited to the ratio of bit width of the number format (2$\times$ from double to single, $4\times$ from double to half). In practice, with accuracy maintained, the realizable speedup has so far been typically less than 2$\times$ even after going to half-precision for some parts of the calculations.
These applications include aerodynamics, astrophysical fluid dynamics, turbulence research, weather and climate.
However, going forward, more innovative techniques such as adaptive precision SpMV \cite{graillat_adaptive_spmv_2024} and co-design of applications with library developers may further improve these speedups, especially by taking advantage of the fact that double precision levels of accuracy are not always required.

The trend of widening gap between double-precision and low-precision arithmetic throughput on hardware platforms is slated to continue to accelerate.
On upcoming hardware platforms, we should keep an eye on the relative ratio of low-precision tensor core compute throughput versus regular FP64 vector throughput. Once this ratio reaches a high-enough number, modern techniques for approximating high-precision matrix-matrix products (eg., Ozaki splitting) may become relevant for scientific applications.
Further, the energy efficiency advantages of low-precision computing make the research and development of mixed-precision algorithms imperative for sustainable computing.

From all of the above, it is our recommendation that the most used scientific applications be analyzed for opportunities to use lower precision with a co-design approach by a multi-disciplinary team of application developers from the relevant science domain, math library developers, and computer scientists.
Such a team is essential to be able to handle the breadth of required tasks: domain specific error analysis and transformations, mixed-precision algorithm development and analysis for recurring motifs, carefully tuned implementation and optimization to ensure effective hardware utilization, as well as analysis tools to pinpoint opportunities and issues on a given hardware platform.
The involvement of math library developers and computers scientists will also help with transferring ideas from one science domain to another, to increase the pay-off in terms of accelerating the ecosystem of scientific applications.

\backmatter

\section*{Declarations}

This article does not contain any studies with human or animal participants.

The authors declare no potential conflicts of interest with respect to the research, authorship, and/or publication of this article.

\subsection*{Funding}
This manuscript has been authored by UT-Battelle, LLC, under contract DE-AC05-00OR22725 with the US Department of Energy (DOE). The US government retains and the publisher, by accepting the article for publication, acknowledges that the US government retains a nonexclusive, paid-up, irrevocable, worldwide license to publish or reproduce the published form of this manuscript, or allow others to do so, for US government purposes. DOE will provide public access to these results of federally sponsored research in accordance with the DOE Public Access Plan (https://www.energy.gov/doe-public-access-plan).

This research used resources of the Oak Ridge Leadership Computing Facility at the Oak Ridge National Laboratory, which is supported by the Office of Science of the U.S. Department of Energy under Contract No. DE-AC05-00OR22725.

%\begin{acks}
%\end{acks}

%\printbibliography
\bibliography{refs}

\end{document}